\documentclass[lettersize,journal]{IEEEtran}

\usepackage{amsmath,graphicx, amssymb, color, subcaption, amsthm, multirow, array, adjustbox, booktabs, lipsum, float, hyperref}
\newtheorem{exmp}{Example}

\newcommand{\rev}[1]{\textcolor{black}{#1}}
\usepackage{cite, soul}

\title{ENN: A Neural Network with DCT Adaptive Activation Functions}

\author{Marc Martinez-Gost$^{1, 2}$ \qquad Ana Pérez-Neira$^{1, 2, 3}$ \qquad Miguel Ángel Lagunas$^{2}$\\
$^{1}$ Centre Tecnològic de Telecomunicacions de Catalunya, Spain \\
$^{2}$ Dept. of Signal Theory and Communications, Universitat Politècnica de Catalunya, Spain\\
$^{3}$ ICREA Acadèmia, Spain
\thanks{This work is part of the project IRENE (PID2020-115323RB-C31), funded by MCIN/AEI/10.13039/501100011033.}
}

\begin{document}
%
\maketitle
\begin{abstract}

The expressiveness of neural networks highly depends on the nature of the activation function, although these are usually assumed predefined and fixed during the training stage. \rev{Under a signal processing perspective}, in this paper we present Expressive Neural Network (ENN), a novel \rev{model} in which the non-linear activation functions are modeled using the Discrete Cosine Transform (DCT) and adapted using backpropagation during training. This parametrization keeps the number of trainable parameters low, is appropriate for gradient-based schemes, and adapts to different learning tasks. This is the first non-linear model for activation functions that relies on a signal processing perspective, providing high flexibility and expressiveness to the network.
We contribute with insights in the explainability of the network at convergence by recovering the concept of bump, this is, the response of each activation function in the output space.
Finally, through exhaustive experiments we show that the model can adapt to classification and regression tasks. The performance of ENN outperforms state of the art benchmarks, providing above a 40\% gap in accuracy in some scenarios.
\end{abstract}

\begin{IEEEkeywords}
Neural networks, adaptive activation functions, discrete cosine transform, explainable machine learning.
\end{IEEEkeywords}

\section{Introduction}
\label{sec:intro}
Function approximation is a fundamental problem across many domains, such as data analysis, control systems and communications. When the explicit function is not available but input-output data pairs are, the function can be revealed by minimizing a criterion loss in a supervised setting.
The problem increases in complexity when the function is non-linear, for which many signal processing techniques have been developed. For instance, least squares \cite{Che91}, orthogonal function approximation \cite{shi96, Pat93}, kernel methods \cite{Man07} and neural networks \cite{Lia16, Yan13, Fer05}, among others. The last decades have suffered an unprecedented growth in the development of artificial neural networks for function approximation due to its empirical success. The inception of neural networks as universal approximators boosted its development across many fields and applications, with different architectures built according to the task and data types to handle. Nevertheless, the expressiveness of the neural network is related to the non-linear activation function, which is usually assumed fixed. An overlooked field of research is in adaptive activation functions (AAF), where not only the weights of the neural networks are trained, but the non-linearities too \cite{Che96}.

In our previous work \cite{Gos23} we introduced the Discrete Cosine Transform (DCT) to approximate an univariate non-linear function in a joint communication and computing setting. Further, in \cite{Per23} we show how a gradient-based algorithm can be used to tune the DCT coefficients to approximate a function in a supervised setting.
However, extending the results to multivariate functions is not trivial: The number of required parameters to approximate the function increases exponentially with the number of input variables, and their corresponding indexes are unknown when the explicit function is not available. In other words, it is cumbersome how the top relevant coefficients can be learnt in a supervised fashion using labeled data.

In this work we propose to extend the capabilities of the DCT with a novel neural network \rev{model} that integrates the DCT to represent and adapt the activation functions. We call this \rev{model} Expressive Neural Network (ENN). We exploit the fact that a 2-layer neural network can theoretically represent any function and expand the representation capabilities of the network by adapting the activation functions at each neuron. The advantage of approximating an univariate function with the DCT is twofold: A small number of coefficients is required due to its high energy compaction, and the approximation error is easily controlled by the magnitude of the disregarded coefficients.
In this work we also show that the DCT coefficients can be learnt using backpropagation in a supervised fashion and in the same pass as the standard network linear weights. In this way, the architecture is no different from a standard feed-forward neural network with fixed activation functions. From the learning perspective, using the DCT to model the activation functions brings the following benefits:
\begin{itemize}
    \item Network size: The number of parameters in the network grows linear with the number of neurons, which is small due to energy compaction of DCT.
    
    \item Backpropagation: The tuning algorithm can be implemented because the DCT coefficients are real and ordered in decreasing magnitude. Besides, there exist analytical closed-form solutions for backpropagation.
    
    \item Gradient behavior: The basis functions (cosines) are real and bounded, which prevents exploding gradients. Likewise, since the Fourier representation creates a periodic function that does not saturate, it also prevents vanishing gradients.
    
    \item Task adaptability: The output non-linearity is automatically adapted depending on the task (e.g., classification or regression) \rev{without changing the loss function}.
\end{itemize}

While we provide a general formulation for multivariate real functions, we constrain the analysis to bivariate functions. This allows to visualize the results and intuitively understand how the network is adapting the activation functions. In this respect, we recover the concept of \textit{bump}, which is the non-linear enclosure that each activation function generates in the output space \cite{Dud01}. The global response of the network corresponds to a weighted sum of all bumps generated at the hidden layer. This concept allows to gain insights in how the network decides to exploit the periodic nature of the DCT model and create the boundaries for classification problems.
In this work we focus on two general problems of function approximation, namely classification and regression. In the former there are two hypothesis associated to a function, this is, $\mathcal{H}_0$ when $f(x_1,x_2)<0$ and $\mathcal{H}_1$ otherwise. In regression the goal is to approximate the function $f(x_1,x_2)$.
\rev{Our primary goal is to show the interpretability and expressiveness of the DCT-based non-linearities in small networks, which is why we do not consider standard datasets and large models. We leave these considerations for future work, along with other learning aspects, such as overfitting}. The source code of this study is openly available on \url{http://github.com/marcmartinezgost/enn}.

The main contributions of this paper are described in the following:
\begin{enumerate}
    \item We define ENN, a novel neural network \rev{model} with non-linear AAF that are parameterized by the DCT. This allows to adapt DCT coefficients in a supervised fashion and learn specific non-linearities according to the task. This results in a highly flexible and expressive \rev{model}.

    \item We develop analytical closed-form expressions to adapt the non-linearities with backpropagation. While we choose the Least Mean Squares (LMS) algorithm to update the network parameters, the architecture remains a feed-forward neural network and any alternative algorithm can be used.

    \item We provide insights in the field of explainable machine learning (XML). We recover the concept of bump, which allows to interpret how the non-linearities are adapted and what is the global response of the neural network. Furthermore, we show how the DCT model helps to dimension the network width, this is, the number of hidden neurons. \rev{Particularly, the network converges to duplicated activation functions with opposed linear weights in the output layer. This is, the network cancels out the information coming from several neurons when the task does not require that many parameters}.

    \item We provide extensive experiments in both classification and regression setups for which ENN outperforms all the benchmarks.
    In classification tasks, ENN outperforms fixed activation functions up to 40\% in accuracy.
    With that we show how the expressiveness of network highly depends on the activation function, without the need of increasing the size of the network.
\end{enumerate}


The remainder of this paper is organized as follows. In Section II, we present a literature review on neural networks for function approximation. Fourier models for non-linear representation are presented in
Section III. We present ENN in Section IV and propose the learning procedure for supervised tasks in Section V. The simulation results are shown in Section VI and we conclude the paper in Section VII.\\

\noindent
\textbf{Notation}: Lowercase and uppercase bold symbols correspond to vectors and matrices, respectively;
$\mathbf{s}[m]$ corresponds to the $m$-th entry of vector $\mathbf{s}$;
$\mathbb{R}$ stands for the set of real numbers and $\nabla$ for the gradient.

\section{Literature Review}

The universal approximation theorem is a well-known result in mathematics stating that a 2-layer neural network can represent any continuous function with an arbitrary number of neurons \cite{Hor91, Les93}. The theory behind neural networks has been further developed, providing bounds on the number of required neurons.

In a different line of research, the Kolmogorov-Arnold (KA) representation theorem shows how a multivariate function can be represented by functions of only univariate functions \cite{Kol57}:
\begin{equation}
    f(x_1,\dots,x_n) = \sum_{i=1}^{2n+1} \Phi_i\left(
    \sum_{j=1}^{n} \phi_{ij}(x_j)
    \right),
    \label{eq:kolmogorov}
\end{equation}
where $\Phi_i$ and $\phi_j$ are termed the outer and inner functions, respectively. This result seems to be tightly connected to a 2-layer neural network, since the inner functions correspond to the hidden layer transformation and the outer functions to the output neuron.  What is more, the inner functions do not depend on the function $f$ to implement, which resembles the activation functions in neural network architectures. However, the formulation is not exactly identical: the KA representation requires $n$ inner functions for each input variable, while a neural network implements a unique function for each linear combination of inputs. In this way, although there is an extensive literature motivating the development of neural networks with the KA theorem \cite{Hec87, Sch21, Nak04}, there are still many gaps to be resolved. What is more, the theorem is not constructive and the functions in \eqref{eq:kolmogorov} are highly non-linear.

Despite its success in many applications, neural networks undergo several shortcomings that limit the interpretability of the results: optimization algorithms are easily trapped in local minima, convergence heavily depends on initialization and fail to converge when high non-linearities exist.
Usually, each neuron in feedforward neural networks implements a linear combination and a non-linear mapping. The former is usually trained, while the latter remains fixed. The non-linear activation function allows to generate non-linear mappings, which increases the expressiveness of the network. A common choice is the sigmoid function, although it exhibits well-known issues in its implementation: since the sigmoid saturates at large magnitudes the propagated gradients vanish, slowing down the learning process.
The rectified linear unit (ReLU) replaced the sigmoid activation function because it does not suffer from vanishing gradients and it
is computationally efficient, which results in faster convergence. Despite its popularity, ReLU also experiences several weaknesses that hinder the learning capacity of the \rev{model}: the corresponding neuron can become dead when the output remains negative for a long time; likewise, the output is unbounded, which may produce the opposite effect of exploding gradient, making the network unstable.

Many variations and novel activation functions have been designed to enhance the performance of neural networks, although this highly depends on the application and the statistics of the data. In this respect, few authors have tackled the problem of AAF, in which the non-linear mapping is also trained to enhance the expressiveness of the network. In \cite{Che96} the saturation level and slope of an S-shaped activation function can be tuned independently, which increases the expressiveness with respect to the sigmoid function. In a different vein, several authors proposed to approximate an arbitrary activation function using cubic splines \cite{Cam96, Vec98}. Later on, some authors introduced an activation function as a weighted sum of basis, such as sine, Gaussian or sigmoid \cite{Xu00, Xu01}. More recently, the authors in \cite{Gom09} propose an asymmetric S-shaped activation function, but only for regression tasks. The issues derived from these perspective is that they either constrain the geometry of the activation function, or the parametrization is too complex. In this work we empirically prove that having pre-specified activation functions result in suboptimal performance.

In the context of deep learning, several AAF have been designed as well. In \cite{Ago14} the authors propose a gradient adaptive piecewise linear activation function as a sum of hinge-shaped functions. Although they theoretically prove that any piecewise linear function can be expressed as so, this constrains the function to be linear at both extremes. In other words, it does not prevent the neural network from exploding gradients. In \cite{Jin16} an S-shaped rectified linear activation function is proposed. While this model can learn non-linear and non-convex representations, it is not smooth and still constrains the activation to be defined in a piece-wise fashion. The authors in \cite{Hou16} develop a piecewise polynomial form which can approximate any continuous function. However, the exponential nature of the polynomials may increase the dynamic range of the parameters, which is a non-desirable property for gradient-based procedures. In \cite{Dan20}, the ReLU is substituted by an S-shaped activation, which slightly outperforms the ReLU and its variants in different deep learning tasks. In \cite{Ram18, Man18, Qia18, Var21} different authors propose to model the activation as a sum of basis functions for deep learning tasks.

\rev{There have been also attempts to implement activation functions using Fourier models. In \cite{Lee22} the authors consider a binarized neural network, where tunable parameters only have 1 bit. This makes the activation function to be the step function and, therefore, the gradient is zero almost everywhere. To make the model trainable, the step function is approximated with a Fourier series, which prevents the gradient vanishing effect.
Other works propose numerical solutions to specific mathematical models by applying Extreme Learning Machines (i.e., feed-forward neural networks) with Fourier basis functions. In \cite{yan22}, the Fourier series basis is used to solve the one-dimensional asset pricing model. In \cite{ma22} the authors extend the work with a product of trigonometric functions to cope with the specific multi-dimensional case that solves the generalized Black–Scholes partial differential equation. In the same vein of ELM with trigonometric activation functions is \cite{zho19}, where numerical solution is found to integro-differential equations for risk theory. However, all these works, which do not resort to gradient-based backpropagation to work, are designed to solve specific numerical system of equations. Conversely, the ENN tries to solve the general problem of multi-dimensional function approximation.}

In the following section we propose a Fourier-based parametrization for non-linear functions.
This is the first approach to model AAF from a signal processing perspective. The proposed model does not constrain the shape of the activation function and keeps the number of trainable parameters low while circumventing the shortcomings of the previously proposed non-linear models.

\section{Fourier Models for Non-linear Functions}
Consider a scalar univariate function $f(x)$ to be approximated over the input variable range $x\in[-1,1]$.
One of the most popular approximations for non-linear systems is the Volterra model. Given the function representation by the inverse Fourier transform,
\begin{equation}
    f(x) = \frac{1}{2\pi} \int_{-\infty}^{\infty} F(w)e^{jwx} \,dw,
    \label{eq:inv_fourier}
\end{equation}
where $w$ is the frequency variable and $F(w)$ are the Fourier coefficients, the Volterra model is obtained by using the Taylor series expansion of the exponential family:
\begin{align}
    f(x) &= \frac{1}{2\pi} \int_{-\infty}^{\infty} F(w)\left(\sum_{n=0}^\infty \frac{(jw)^n}{n!}x^n\right) \,dw\nonumber\\
    &= \sum_{n=0}^\infty\underbrace{\left( \frac{1}{2\pi}\frac{1}{n!} \int_{-\infty}^{\infty} F(w)(jw)^n \,dw \right)}_{\textstyle c_n\mathstrut} x^n
    \label{eq:volterra}
\end{align}

The building blocks of \eqref{eq:volterra} are power functions of the input variable, which generate a polynomial approximation. Notice that the coefficients $c_n$ correspond to the $n$-th derivative at the origin divided by the factorial of the index. Volterra has not been widely used in practice because the exponential nature of the basis functions heavily increase the dynamic range, even when the input is bounded. Moreover, outside the dynamic range the approximation is unbounded and cannot be controlled. This happens because Volterra is a Taylor approximation and it presents high accuracy near the origin only. \rev{Moreover, the coefficients change with respect to how many are preserved, as the kernels are not orthogonal.}
All these concerns make this approximation not suitable for gradient-based learning algorithms \cite{Per23}.

One possible approach to solve these issues is the representation using a finite number of terms in the Fourier approximation in \eqref{eq:inv_fourier}, which corresponds to the Discrete Fourier Transform (DFT). Outside the dynamic range of $x$ the function is periodically extended, generating discontinuities at the border. As a result, the number of coefficients required to approximate the function is highly sensitive to these discontinuities.
To prevent this phenomenon the function can be extended with even symmetry and, then, periodically extended. This smooths out the edges and its corresponding derivatives, reducing the number of required coefficients with respect to the DFT. This is known as the Discrete Cosine Transform (DCT), which has the following expression:
\begin{equation}
f(x) \approx \sum_{q=1}^{Q} g_qF_q \cos\left(\frac{\pi (q-1)(2z+1)}{2N}\right),
\label{eq:idct}
\end{equation}
with $z=\frac{N}{2}(1+x)$, $g_1=1/\sqrt N$ and $g_q=\sqrt{2/N}$ otherwise. The $F_q\in\mathbb{R}$ are termed the DCT coefficients.
Notice that for functions with odd symmetry, only the odd coefficients are retained. In general, the quality of approximation is more than sufficient for $Q=12$ (i.e., 6 coefficients in odd functions). For the sake of brevity, the following definition will be used when needed:
\begin{equation}
    cos_i(x) = \cos\left(
\frac{\pi}{2N}\left(2i-1\right)\left(N(x+1)+1
\right)
\right)
\end{equation}

In \cite{Per23} we propose an adaptive design in which the DCT coefficients of \eqref{eq:idct} are tuned using the LMS algorithm in a supervised setting to approximate an univariate function. There are plenty of advantages in using the DCT representation: The required coefficients to provide the same quality (i.e., approximation error) are fewer and these are real and ordered in decreasing magnitude. 
Since the basis functions are orthogonal, the approximation error can be easily controlled by the magnitude of the disregarded coefficients, and
it simplifies the convergence of the learning procedure.
Furthermore, see that the coefficient index appears in the phase of \eqref{eq:idct}, so the approximation is real and bounded, even when the input exceeds the dynamic range. All these features make the DCT an appropriate function approximation, whose  coefficients can be learnt by a gradient-based rule.

\subsection{Extension of the DCT to multiple input variables}
\label{sec:2-dct}
The extension of the DCT representation to multiple input variables is not trivial. Consider a bivariate function, which could be parameterized by a 2-dimensional (2D) DCT as
\begin{align}
    f(x_1,x_2) = \sum_{n=1}^{N}\sum_{m=1}^{N} F_{nm}\cos_n(x_1)\cos_m(x_2)\nonumber\\
    \approx
    \sum_{n=1}^{Q}\sum_{m=1}^{Q} F_{nm}\cos_n(x_1)\cos_m(x_2)
\label{eq:2-dct}
\end{align}

Note that this \rev{model} has the following drawbacks:
In general the number of coefficients grows exponentially with the number of inputs, $M_0$, as $Q^{M_0}$, as well as the complexity of the DCT; while the coefficients in the DCT are ordered in decreasing magnitude, the structure of the indexes is broken in the 2D-DCT. This is, the location $(n,m)$ of the relevant coefficients changes with the function of interest; this implies that, unless the function is known a priori, the indexes $(n,m)$ are unknown and a supervised learning procedure is hard to implement. This happens for instance in classification problems, where the function is unknown and to be discovered.

In the following we will present ENN, a neural network \rev{model} integrating the DCT in a single dimension and whose coefficients can be trained in a supervised fashion.

\section{DCT-Adaptive Activation Functions}
A perceptron (or neuron) is a non-linear processor involving a weighted sum and a non-linear function. It is described by the following expressions:
\begin{align}
    z &= a_0 + \sum_{i=1}^{M_0} a_ix_i=\mathbf{a}^T[1\,\, \mathbf{x}^T]^T
    \label{eq:linear}\\
    \overline{z} &= \frac{N}{2}(z+1)
    \label{eq:normalization}\\
    \hat{y} &= \sigma(\overline{z})\approx
    \sum_{q=1}^{Q/2} F_q\cos\left(\frac{\pi (2q-1)(2\overline{z}+1)}{2N}\right)\nonumber\\
    &\qquad\quad\,=\sum_{q=1}^{Q/2} F_q\cos_q(z)
    \label{eq:non_linearity}
\end{align}
In \eqref{eq:linear}, $\mathbf{x}$ contains $M_0$ inputs and a 1 is appended for the bias term $a_0$. The normalization in \eqref{eq:normalization} is needed to map the input to $[0,N]$, assuming the input is confined to $[-1,1]$. Nevertheless, as it will be seen later, the first linear transformation may map $\overline{z}$ to a different range, providing expressiveness to the network. Finally, the non-linear activation function $\sigma(\cdot)$ is approximated by the DCT with $Q/2$ coefficients. Without loss of generality, $g_q$ is assumed to be integrated in the DCT coefficient $F_q$. In \eqref{eq:non_linearity} we assume the non-linearity to have odd symmetry, so that only the odd coefficients are retained. As it will be explained later on in Sec. \ref{sec:expressiveness} and shown empirically in Sec. \ref{sec:results}, this does not prevent the network from learning only odd activation functions.
A total of $M_0+Q/2+1$ parameters per perceptron are to be trained, which only represents an increment of $Q/2$ coefficients with respect to a standard perceptron with a fixed activation functions. As mentioned in the previous section, $Q$ is small due to the energy compaction property of the DCT (e.g., around $Q/2=6$ coefficients).


\begin{exmp}[Linear discriminant]
\label{exmp:linear_discriminant}
Assume $M_0=2$, we want to discriminate when $x_1>x_2$. This can be framed as a classification problem, for which we can take $\mathbf{a}=a[0\,1\,-1]^T$, where $a$ is a positive constant. Then, any odd monotone increasing function will discriminate the two hypothesis. For the sake of simplicity, take $a=1$ and the non-linearity approximated with only one coefficient (i.e., $F_1=-1$). This results in
\begin{align}
    z &= x_1-x_2
    \label{eq:linear_example}\\
    \overline{z} &= \frac{N}{2}(x_1-x_2+1)
    \label{eq:normalization_example}\\
    \hat{y} &= -\cos\left(\frac{\pi (2\overline{z}+1)}{2N}\right)\approx
    \sin\left(\frac{\pi}{2}(x_1-x_2)\right)
    \label{eq:non_linearity_example}
\end{align}
The solution in \eqref{eq:non_linearity_example} provides a soft-decision, whereas a hard-decision would take the sign, i.e., $\text{sign}(\hat{y})$.
While there are infinite solutions for this example, the proposed scheme only increases the complexity by one extra coefficient (i.e., $F_1$). Implementing a linear $\sigma(\cdot)$ would require many coefficients and provide no further benefit in terms of performance.
\end{exmp}

In Example \ref{exmp:linear_discriminant} there are infinite minimums, all optimal. Nevertheless, in more complex problems (e.g., high-order discriminants) the expressiveness of a single perceptron is not enough. In light of the universal approximation theorem, in the following we will increase the capabilities of the network by including a hidden layer of several neurons. This \rev{model}, termed ENN, will be fully-adaptive as both the linear weights and the activation functions will be trained in a supervised fashion.

\subsection{Expressive Neural Network (ENN)}

\begin{figure}[t]
    \centering
    \includegraphics[width=\columnwidth]{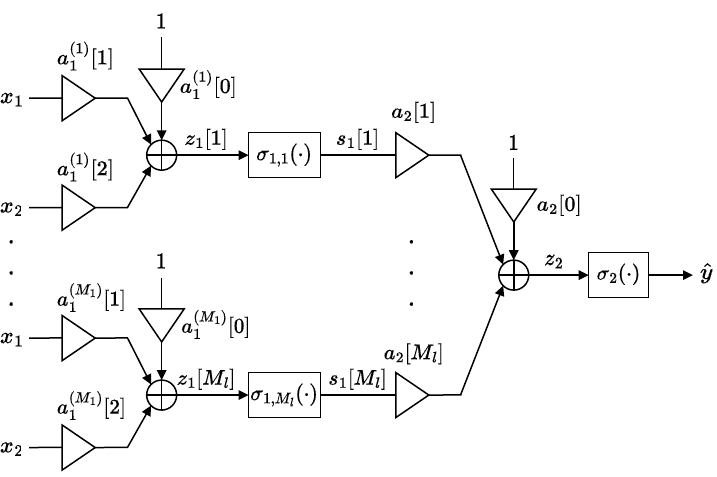}
  \caption{A 2-layer perceptron with $M_1$ neurons in the hidden layer.}
  \label{fig:2_layer_perceptron}
  \vspace{-0pt}
\end{figure}

As a single layer perceptron limits the capabilities of the network, we will include another processing layer to increase its expressiveness. For a general multi-layer perceptron of $L$ layers, the following expressions show the perceptron signals at the $l$-th layer:
\begin{align}
    \mathbf{z}_l &= \mathbf{A}_l^T[1\,\, \mathbf{s}_{l-1}^T]^T
    \label{eq:linear_multi}\\
    \overline{\mathbf{z}}_l &= \frac{N}{2}(\mathbf{z}_l+1)
    \label{eq:normalization_multi}\\
    \mathbf{s}_l &= \sigma_{l,m} (\overline{\mathbf{z}}_l)
    \label{eq:non_linearity_multi}
\end{align}
for $l=1,\dots,L$. The first input and last output correspond to $\mathbf{s}_{0}=\mathbf{x}$ and $\hat{y}=\mathbf{s}_L$, respectively. The matrix of linear weights is $\mathbf{A}_l=[\mathbf{a}_l^{(1)}\,\dots\,\mathbf{a}_l^{(M_l)}]$ and $M_l$ stands for the number of perceptrons at layer $l$, with $M_0$ being the number of inputs.
The operations in \eqref{eq:normalization_multi} and \eqref{eq:non_linearity_multi} are performed element-wise. The AAF in the ENN, this is, the $m$-th element of \eqref{eq:non_linearity_multi} is computed as
\begin{equation}
    \mathbf{s}_l[m] =
    \sum_{q=1}^{Q/2} F_{q,l}^{(m)}\cos_q(\mathbf{z}_l[m]),
    \label{eq:non_linearity_neuron}
\end{equation}
where $F_{q,l}^{(m)}$ corresponds to the $q$-th coefficient of the $m$-th perceptron at the $l$-th layer. As explicitly shown in \eqref{eq:non_linearity_multi} and  \eqref{eq:non_linearity_neuron} the activation function is not necessarily the same at each neuron, although the number of DCT coefficients $Q$ is kept constant for the whole network. The last activation function could be substituted by an step function in the classification setup and a linear for regression. However, this will be maintained so that the network can self-adapt depending on the nature of the problem.

\begin{figure}[t]
    \centering
    \vspace{15pt}
    \includegraphics[width=\columnwidth]{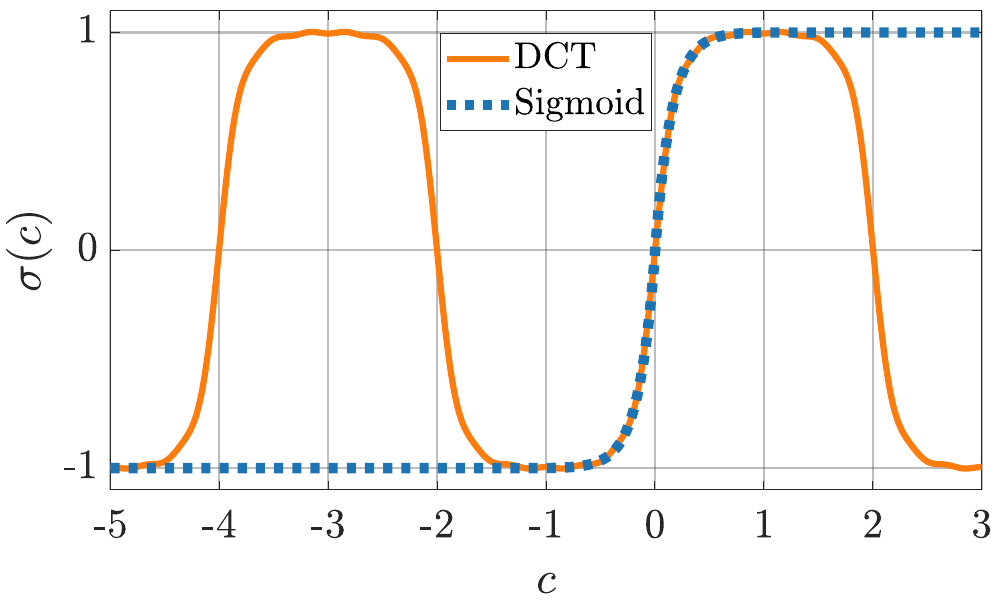}
  \caption{Sigmoid function (dashed) and its DCT representation in the $[-1,1]$ range (solid).}
  \label{fig:sim_DCT_sigmoid}
\end{figure}

As opposed to the $M_0$-dimensional DCT in Sec. \ref{sec:2-dct}, the proposed \rev{model} does not require implementing the DCT in $M_0$ dimensions and can be trained in a supervised fashion. Furthermore, the number of coefficients grows linearly as $M_0Q$. Throughout the rest of the paper we assume a two input variable, namely $M_0=2$, and $L=2$ layers. These layers are sequentially termed hidden and output layer. Fig. \ref{fig:2_layer_perceptron} shows the architecture of the two-layer perceptron with $M_1$ neurons in the hidden layer and expression \eqref{eq:input-output_2_layers} shows the input-output relationship.

\begin{figure*}[t]
\normalsize
\setcounter{equation}{16}
\begin{align}
\hat{y} =&\sum_{m=1}^{Q/2} F_{m,2}^{(1)}\cos_m\left( 
a_2^{(1)}[0] + \sum_{k=1}^{M_1}a_2^{(1)}[k]
\sum_{q=1}^{Q/2} F_{q,1}^{(k)}\cos_q\left(
a_1^{(k)}[0] + a_1^{(k)}[1]x_1 + a_1^{(k)}[2]x_2
\right)
\right)
\label{eq:input-output_2_layers}
\end{align}
\hrulefill
\vspace*{4pt}
\end{figure*}

\subsection{Expressiveness of periodic activation functions}
\label{sec:expressiveness}

As it will be shown in the experimental results, the expressiveness of the DCT comes from its periodic nature. Fig. \ref{fig:sim_DCT_sigmoid} shows the sigmoid function and the corresponding DCT representation in the $[-1,1]$ range using $Q/2=6$ coefficients. While the sigmoid function saturates outside the range, the DCT approximation offers a periodic infinite non-linearity. This offers more capacity to the activation function, as the network may choose to work at an increasing range (e.g., $[-1,0]$), decreasing range (e.g., $[-3,-1]$), bump range (e.g., $[-1,3]$), valley range (e.g., $[-3,1]$), or even with several periods simultaneously (e.g., $[-5,3]$).
Imposing odd symmetry in the DCT representation does not constrain the resulting activation function to be odd. This Fourier model for non-linearities is the first one to provide such flexibility to activation functions, which is not possible with a fixed non-linearity not implemented with the DCT.

\begin{figure}[t]
\centering
     \begin{subfigure}[b]{0.3\columnwidth}
         \includegraphics[width=\columnwidth]{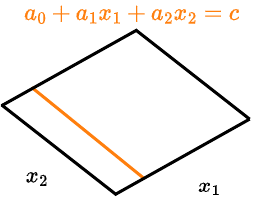}
         \caption[]{Input space.} 
         \label{fig:input_space}
     \end{subfigure}
     \begin{subfigure}[b]{0.30\columnwidth}
         \includegraphics[width=\columnwidth]{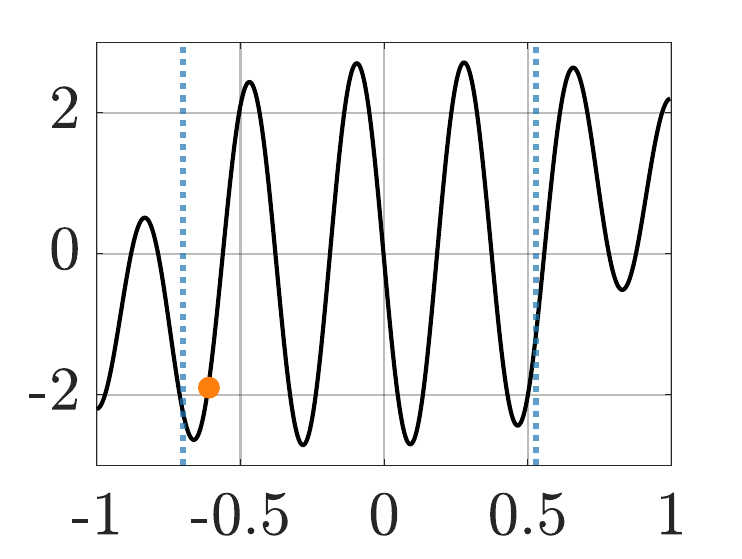}
         \caption[]{Function.} 
         \label{fig:act_dct}
     \end{subfigure}
     \hspace{0pt}
     \begin{subfigure}[b]{0.36\columnwidth}
         \includegraphics[width=\columnwidth]{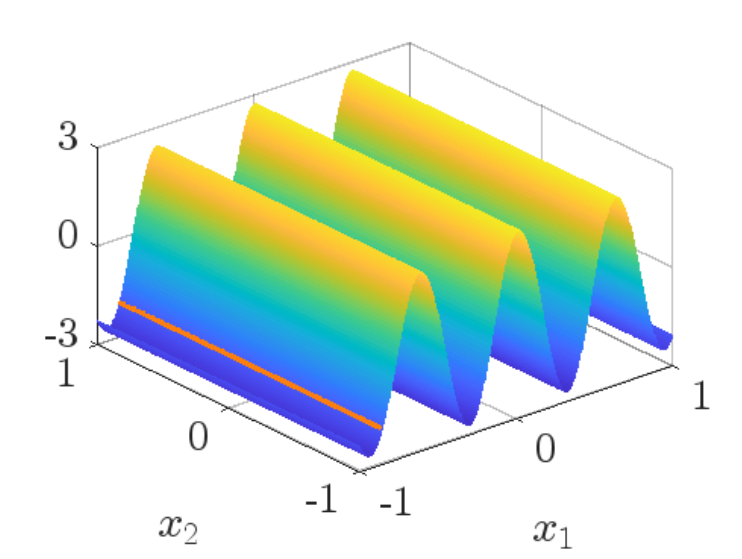}
         \caption[]{Bump.} 
         \label{fig:bump_dct}
     \end{subfigure}

     \caption{\rev{Bump generation. Mapping of a linear combination of two inputs over the response of a non-linear function.}}
     \label{fig:how_to_bumps}
\end{figure}

To understand how the non-linear activation function maps the input to the output space, we use the concept of \textit{bump}. Consider a single neuron with $M_0=2$. Notice that all input pairs satisfying the following equality are mapped to the same non-linear output $\sigma(c)$:
\begin{equation}
    a_0 + a_1x_1 + a_2x_2=c,
    \label{eq: linear_equality}
\end{equation}
where $c$ is a constant value. \rev{Fig. \ref{fig:input_space} shows how expression \eqref{eq: linear_equality} corresponds to a line in the input space. 
Fig. \ref{fig:act_dct} shows the mapping $\sigma(c)$ along with the limits of the function in dashed lines, that correspond to extreme values of $c$. Fig. \ref{fig:bump_dct} shows how all input pairs with $\sigma(c)$ are mapped in the output space. The bump is generated by evaluating all the input data pairs through the linear transformation and the non-linear mapping.}

\section{Supervised Learning}

\subsection{Backpropagation rules}
Consider a dataset $\mathcal{D}$, consisting of $(\mathbf{x}_i,y_i)\in \mathcal{D}$, where the former is a $M_0$-sized vector training sample and the latter is the associated reference (i.e., label in classification, or function value in neuromorphic computing).
Given $y_i$ and $\hat{y_i}$, the error can be computed and propagated throughout the network to adjust the learnable parameters. We assume the loss to be the mean squared error (MSE),
\begin{equation}
    \varepsilon^2=(y_i-\hat{y}_i(\mathbf{x}_i))^2,
    \label{eq:loss}
\end{equation}
for both classification and regression problems. 
Notice in \eqref{eq:loss} we explicitly write the output as a function of the input data. Given a learnable parameter $w$, the chosen algorithm to update it is LMS. The update rule at a given iteration corresponds to
\begin{equation}
    w \leftarrow
    w - \mu\nabla\varepsilon
    =w - \mu\varepsilon\left(-
    \frac{\partial\tilde{y}}{\partial w}
    \right),
\end{equation}
this is, the parameter is updated by the product of the error by the instantaneous gradient of the output with respect to the parameter. The hyperparameter $\mu$ is the step size, which controls the convergence speed of LMS.
We choose the simplest algorithm to learn the parameters in the neural network because the focus of this work is on the \rev{model}, not on the training algorithm. In this respect, the experimental results show that even using the instantaneous gradient, LMS converges to a minimum and outperforms state-of-the-art models. Thus, the LMS may be replaced by any other algorithm to speed up convergence, which is out of the scope of this work. Due to the same reason, we do not include the iteration index in the parameter updates.

For the 2-layer architecture in Fig. \ref{fig:2_layer_perceptron}, there are 2 set of parameters per layer that need to be updated, namely, the DCT coefficients and the weights from the linear transformation.
Starting at the output, the DCT coefficients of the layer $l=2$ are updated as
\begin{equation}
    F_{m,2}\leftarrow
    F_{m,2} + \mu\varepsilon
    \frac{\partial\tilde{y}}{\partial F_{m,2}}=
    F_{m,2} + \frac{4\alpha_1}{Q}\varepsilon
    \cos_m\left(z_2\right),
    \label{eq:LMS_DCT_2}
\end{equation}
for $m=1,\dots,Q/2$. The superscript in the parameter is omitted because there is only one perceptron at the output. In general, the step size is modeled as twice the mismatch $\alpha_1$ divided by the power of the corresponding input. Since the $Q/2$ cosines used to approximate the non-linearity are orthonormal, the total power is $Q/2$. This exhibits another advantage of this parametrization as the power remains always constant.

In order to update the linear weights from the output layer, the backpropagation procedure allows to propagate the error across the network with respect to previously computed derivatives, which makes it a very efficient algorithm. Therefore, these are updates as
\begin{align}
    a_2[k]\leftarrow&\
    a_2[k] + \mu\varepsilon
    \frac{\partial\tilde{y}}{\partial a_2[k]}=
    a_2[k] + \mu\varepsilon
    \frac{\partial\tilde{y}}{\partial z_2}\frac{\partial\tilde{z_2}}{\partial a_2[k]}\nonumber\\
    =&\ 
    a_2[k] - \frac{2\alpha_3}{P_1}\varepsilon
    \frac{\pi}{2}s_1[k]\sum_{m=1}^{Q/2}F_{m,2}(2m-1)\cos_m(z_2),
    \label{eq:LMS_lin_2}
\end{align}
for $k=0,\dots,M_1$, and $P_1=\mathbf{s}_1^T\mathbf{s}_1$. The superscript has also been suppressed in this case. Accordingly, the DCT coefficients in the first layer are updated as
\begin{align}
    F_{q,1}^{(k)}\leftarrow&\
    F_{q,1}^{(k)} + \mu\varepsilon
    \frac{\partial\tilde{y}} {\partial F_{q,1}^{(k)}}=
    F_{q,1}^{(k)} + \mu\varepsilon
    \frac{\partial\tilde{y}}{\partial z_2} \frac{\partial z_2}{\partial s_1[k]} \frac{\partial s_1[k]}{\partial F_{q,1}^{(k)}}
    \nonumber\\
    =&\
    F_{q,1}^{(k)} -\nonumber \\
    &\frac{4\alpha}{Q}\varepsilon
    \frac{\pi}{2} a_2[k]\cos_q(z_1[k])\sum_{m=1}^{Q/2}F_{m,2}(2m-1)\sin_m(z_2),
    \label{eq:LMS_DCT_1}
\end{align}
for $q=1,\dots,Q/2$ and $k=1,\dots,M_1$. And the linear combination of the first layer is updates as
\begin{align}
    a_1^{(k)}[m]\leftarrow&\
    a_1^{(k)}[m] + \mu\varepsilon
    \frac{\partial\tilde{y}} {\partial a_1^{(k)}[m]}\nonumber\\
    =&\
    a_1^{(k)}[m] + \mu\varepsilon
    \frac{\partial\tilde{y}}{\partial z_2} \frac{\partial z_2}{\partial s_1[k]} \frac{\partial s_1[k]}{\partial z_1[k]}
    \frac{{\partial z_1[k]}}{\partial a_1^{(k)}[m]}
    \nonumber\\
    =&\
    a_1^{(k)}[m] +\nonumber \\
    &\frac{2\alpha_4}{P_0}\varepsilon
    \frac{\pi^2}{4} a_2[k]s_0[m]
    \sum_{p=1}^{Q/2}F_{p,2}(2p-1)\sin_p(z_2) \nonumber\\&
    \sum_{q=1}^{Q/2}F_{q,1}^ {(k)}(2q-1)\sin_q(z_1[k]),
    \label{eq:LMS_lin_1}
\end{align}
for $m=1,\dots,M_0$, $k=1,\dots,M_1$, and $P_0=\mathbf{s}_0^T\mathbf{s}_0$. Without loss of generality, $s_0[0]=s_1[0]=1$, which correspond to the bias terms in the linear combination at each layer. Notice that in all the update rules, the derivative exists and inherits the periodic nature of the cosine transform.

To enhance the stability of the convergence, the power of each input, namely $P_0$ and $P_1$, can be computed with a damping effect. For instance,
\begin{equation}
    P_0 \leftarrow \beta P_o + 
    (1-\beta) \mathbf{s}_0^T\mathbf{s}_0
    \label{eq: dampening}
\end{equation}
at each iteration. The $\alpha$ parameter is set smaller in the output layer, which intuitively reduces the propagation of artifacts to previous layers during training.

\subsection{Initialization and Training Procedure}
\label{sec:training}

\rev{The ENN trains all parameters at every backpropagation step, this is, both AAF and linear weights.
The AAF are initialized as linear functions, which has shown to provide the best performance.}
In the hidden layer, the linear weights are initialized to generate bump diversity. 
As seen in Fig. \ref{fig:how_to_bumps}, linear  weights can be interpreted as the orientation of the activation functions (or bump) in the output space. For instance, with $M_1=4$ the linear weights are initialized to $[0,0,1]$, $[0,1,0]$, $[0,1,-1]$ and $[0,-1,1]$. \rev{This
helps in practice to start the learning process with bumps in different orientations.}
Finally, the output layer is initialized with linear weights in a uniform distribution in $[-0.5,0.5]$. This allows to constrain the dynamics and shows that the convergence does not depend on the initialization.

Regarding the LMS setup, for the non-linearities we set $\alpha_1=10^{-3}$ in the hidden layer and $\alpha_2=10^{-4}$ in the output layer. For the linear weights we set $\alpha_3=5\cdot10^{-3}$ in the hidden layer and $\alpha_4=5\cdot10^{-5}$ in the output layer. We keep the step-size constant for all neurons in the same layer. The damping parameter is set to $\beta=0.999$ for both layers. 
\rev{Under the assumption that the error between parameters is uncorrelated, this LMS configuration guarantees that
\begin{equation}
    6\alpha_1+6\alpha_2+\alpha_3+\alpha_4\approx 0.01< 0.1,
\end{equation}
where the right hand side is an upper bound on the error at convergence. Notice that the convergence speed at the output layer is lower than at the hidden layer, which grants an adequate gradient propagation.}


\section{Experimental results}
\label{sec:results}
In the following we will test the ENN for both binary classification and regression problems.
As benchmarks, we will test also the following models, in which they all have the same architecture and differ in the activation functions:
\begin{itemize}
    \item \textbf{ReLU}: it uses a fixed non-trainable ReLU, this is, $\sigma(z)=\max{\{0,z\}}$.

    \item \textbf{Sigm}: it uses a fixed non-trainable sigmoid activation function. Notice that the output saturates for an input of large magnitude.

    \item \textbf{F-DCT:} it uses  a fixed non-trainable sigmoid activation function, but modeled with the DCT. Thus, the function does not saturate and it is periodic.
\end{itemize}

Although not an odd function, we include the ReLU because it is the standard activation function in current neural networks. The Sigm model is also included to compare it with F-DCT and assess the gains of a periodic activation function. 
\rev{We acknowledge that the benchmarks have significantly less parameters than the ENN and they will exhibit lower learning capabilities. However, the scope of this work is to show the expressiveness through the adaptability of non-linear functions instead of increasing the width of the architecture. We constrain the number of neurons because we work on interpretability and want to provide a signal processing perspective to the problem.}

All \rev{models} are built with $M_1=6$ neurons in the hidden layer, and ENN with $Q=12$ parameters in all the non-linearities and $N=512$ samples. However, recall that only $Q/2$ coefficients are different from zero because we impose odd symmetry. These are initialized to approximate the identity function.
Notice that the output activation function in ReLU and Sigm has to be specifically selected to be either sigmoid in classification or linear (i.e., identity) in regression. Conversely, the ENN will automatically adapt it to approach the required function, which is an advantage of this \rev{model} with respect to the benchmarks.

For both classification and regression problems,
a  synthetic dataset  is generated. All samples come from independent uniform distributions in the $[-1,1]$ range for each input variable.
The train and test sets contain 800.000 and 50.000 samples, respectively.

\subsection{Classification}
Table \ref{tab:classification_results} shows the different classification problems that have been evaluated. It displays the ideal decision map along with the order of the discriminant. The metric chosen to compare the different models is the test accuracy, this is, the percentage of correctly classified samples in the test set.

\begin{table}[t]
  \begin{center}
    \begin{tabular}{ccc|cccc}
      \toprule 
      Nº & Map & Order & ReLU & Sigm & F-DCT & ENN\\
      \midrule
      
      
      (P1)&
      \adjustbox{valign=c}{\includegraphics[width=0.3in, angle=180]{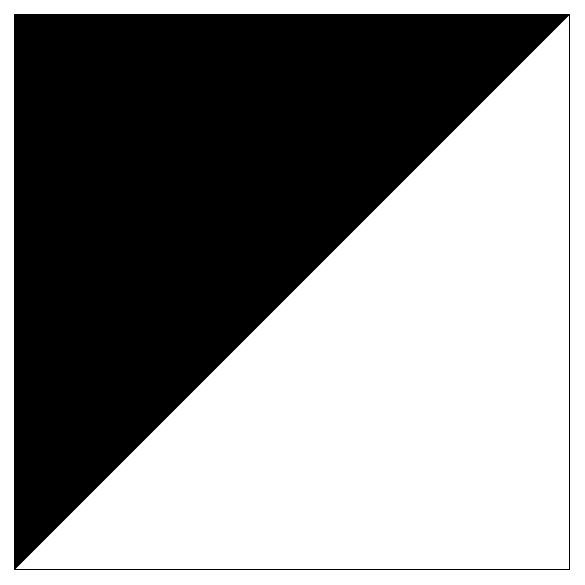}}
      & 1 &
      99.97\% & 99.90\% & 99.84\% & 99.85\% \\
      \midrule
      
      
      (P2)&
      \adjustbox{valign=c}{\includegraphics[width=0.3in, angle=-90]{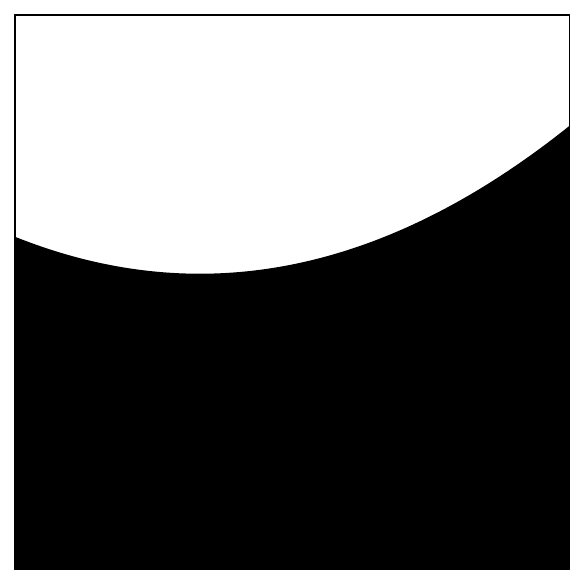}}
      & 2 &
      97.62\% & 97.82\% & 97.43\% & \textbf{99.85\%} \\
      \midrule
      (P3)&
      \adjustbox{valign=c}{\includegraphics[width=0.3in, angle=-90]{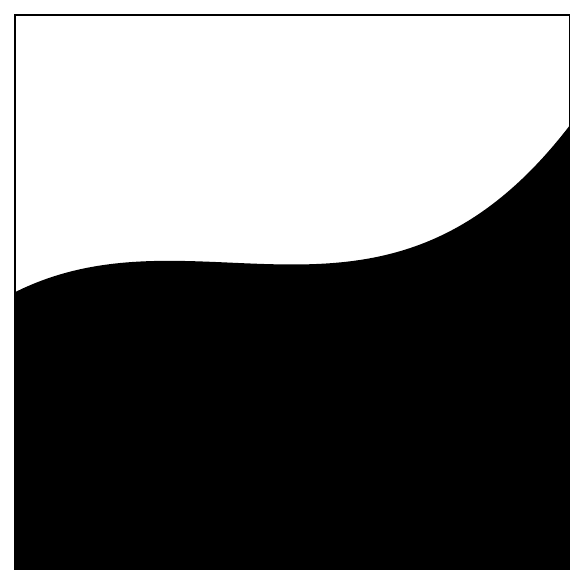}}
      & 3 &
      96.70\% & 96.74\% & 96.17\% & \textbf{99.90\%} \\
      \midrule
      (P4)&
      \adjustbox{valign=c}{\includegraphics[width=0.3in]{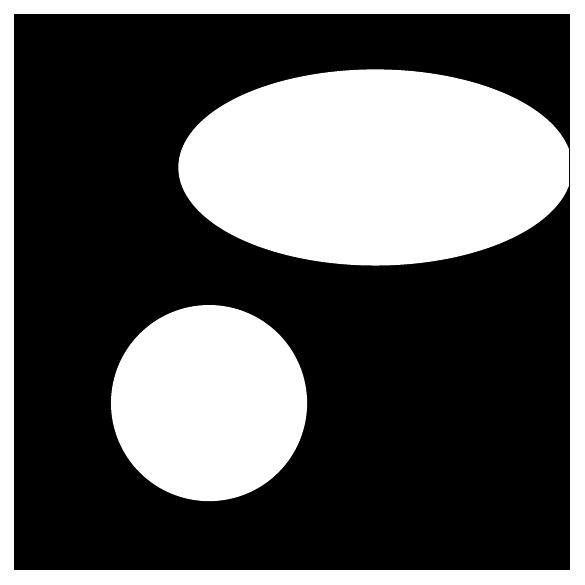}}
      & \multirow{10.5}{*}{$>3$} &
      95.86\% & 96.00\% & 87.36\% & \textbf{99.16\%} \\
      (P5)&
      \adjustbox{valign=c}{\includegraphics[width=0.3in]{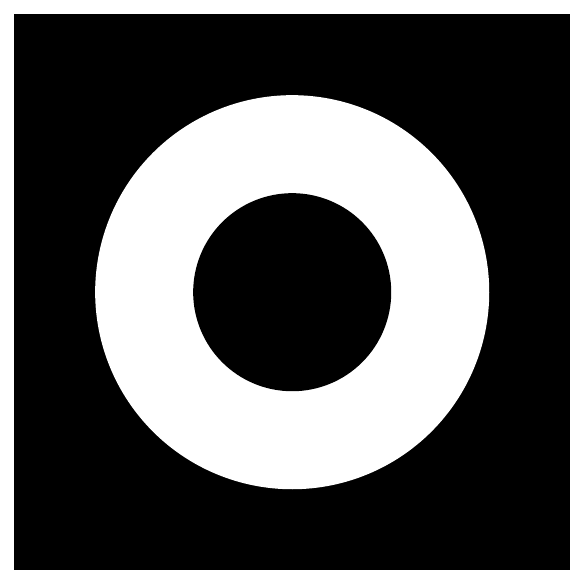}}
      &  &
      91.76\% & 89.68\% & 89.87\% & \textbf{99.48\%} \\
      (P6)&
      \adjustbox{valign=c}{\includegraphics[width=0.3in]{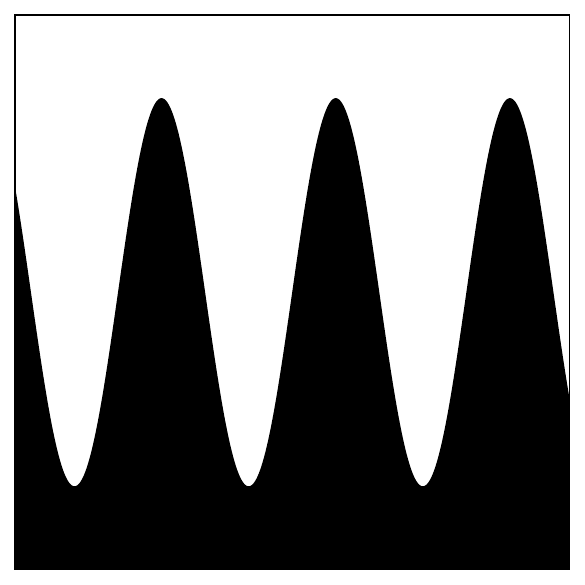}}
      & &
      75.72\% & 75.20\% & 75.20\% & \textbf{99.19\%} \\
      (P7)&
      \adjustbox{valign=c}{\includegraphics[width=0.3in]{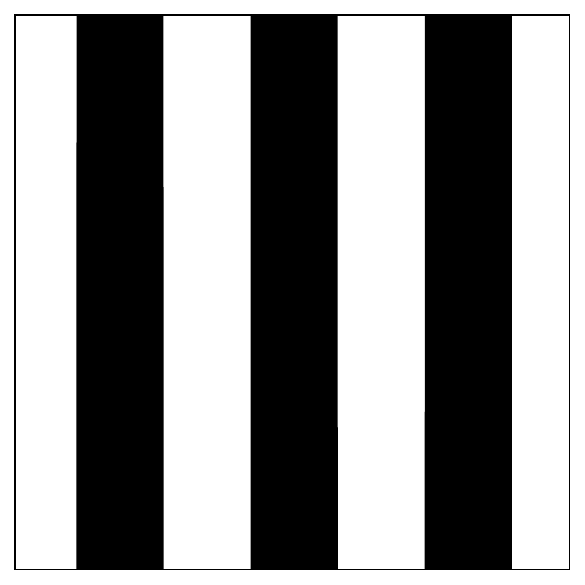}}
      &  &
      51.93\% & 54.56\% & 54.22\% & \textbf{99.62\%} \\
      (P8)&
      \adjustbox{valign=c}{\includegraphics[width=0.285in, angle=-90]{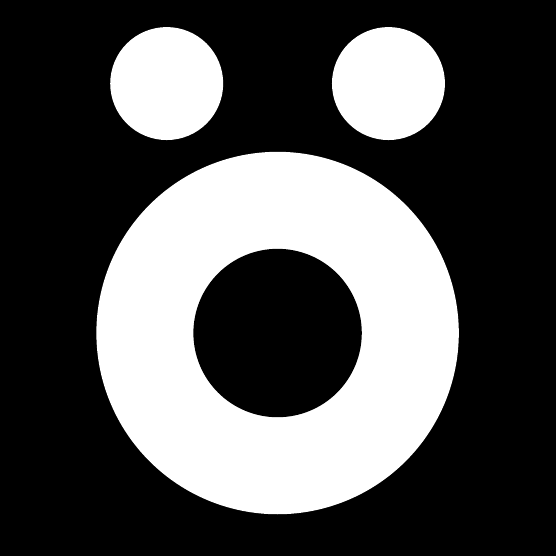}}
      &  &
      75.68\% & 80.16\% & 79.32\% & \textbf{98.22\%} \\
      \bottomrule
    \end{tabular}
    \caption{Accuracy for different binary classification problems. For each case, the ideal map and the order of the discriminant are shown.}
    \label{tab:classification_results}
    \vspace{-10pt}
  \end{center}
\end{table}

As expected, the complexity of the problem increases with the order of the discriminant. As seen in Example \ref{exmp:linear_discriminant}, a single neuron is sufficient to implement a linear discriminant, which results in an excellent performance for all models in (P1). With no surprise, for quadratic (P2) and cubic (P3) discriminants all models are capable of finding the appropriate boundary. However, ENN attains almost an excellent performance. This may be claimed due to the larger number of parameters in the proposed model with respect to the benchmarks.
When moving to high-order classification problems is when the non-adaptive models are not capable of providing a satisfactory solution. Conversely, the ENN model manages to find a local minimum with high accuracy and with very consistent results for a wide variety of problems.

\begin{figure}[t]
\centering
     \begin{subfigure}[b]{0.32\columnwidth}
         \includegraphics[width=\columnwidth]{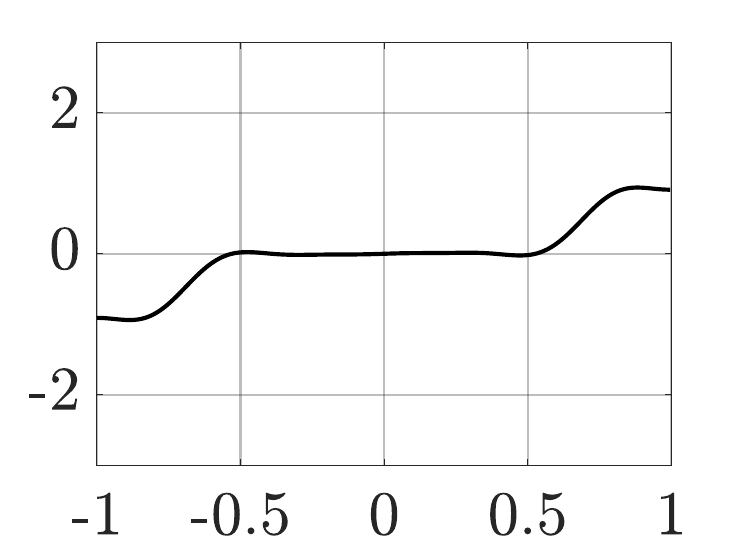}
         \caption[]{Hidden 1.} 
         \label{fig:act_lines_hidden1}
     \end{subfigure}
     \hspace{0pt}
     \begin{subfigure}[b]{0.32\columnwidth}
         \includegraphics[width=\columnwidth]{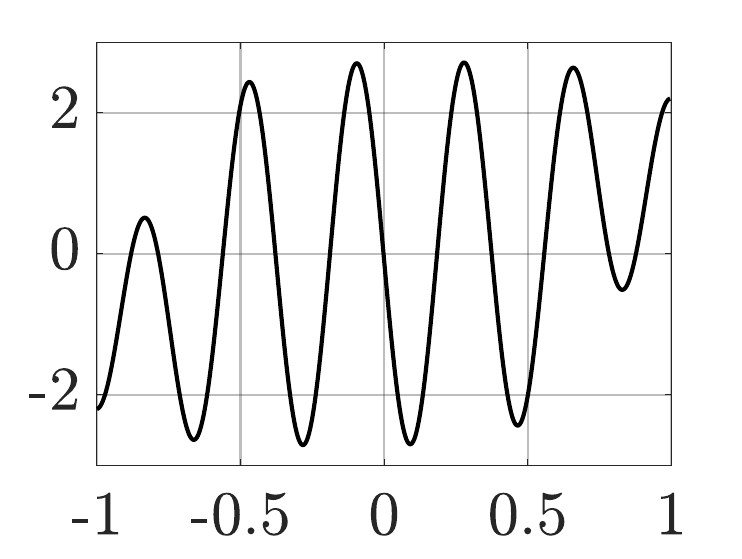}
         \caption[]{Hidden 2.} 
         \label{fig:act_lines_hidden2}
     \end{subfigure}
     \hspace{0pt}
     \begin{subfigure}[b]{0.32\columnwidth}
         \includegraphics[width=\columnwidth]{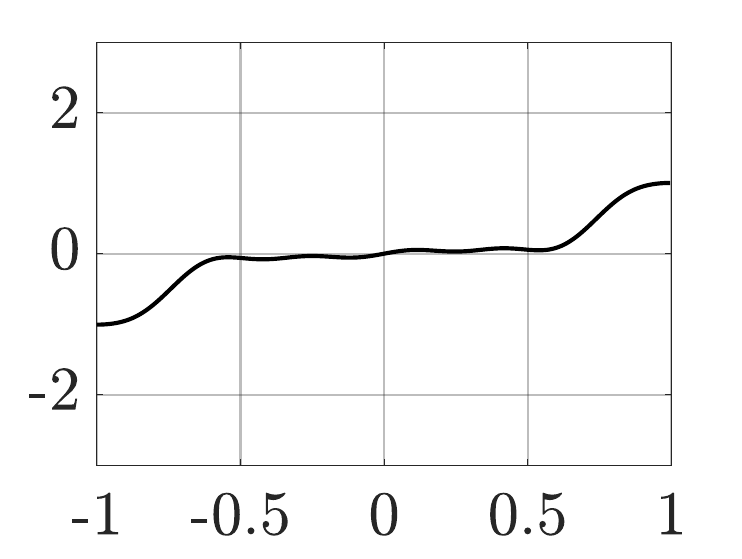}
         \caption[]{Hidden 3.} 
         \label{fig:act_lines_hidden3}
     \end{subfigure}

     \vskip\baselineskip
     \begin{subfigure}[b]{0.32\columnwidth}
         \includegraphics[width=\textwidth]{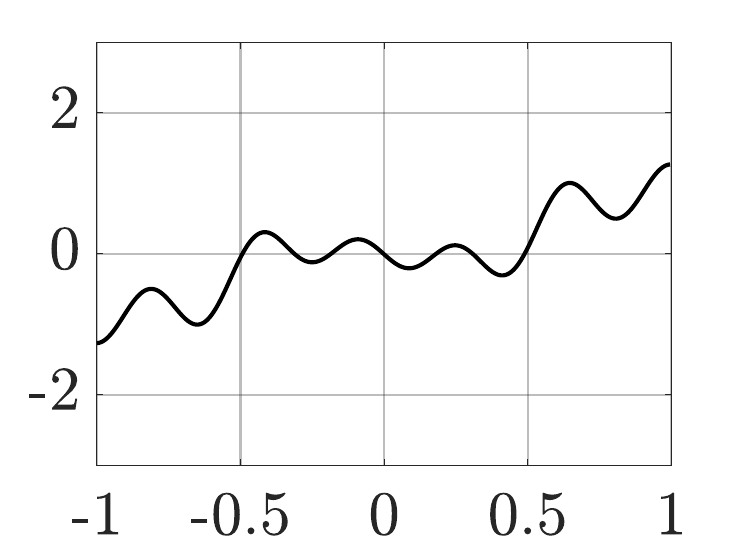}
         \caption[]{Hidden 4.} 
         \label{fig:act_lines_hidden4}
     \end{subfigure}
     \hspace{0pt}
     \begin{subfigure}[b]{0.32\columnwidth}
         \centering
         \includegraphics[width=\columnwidth]{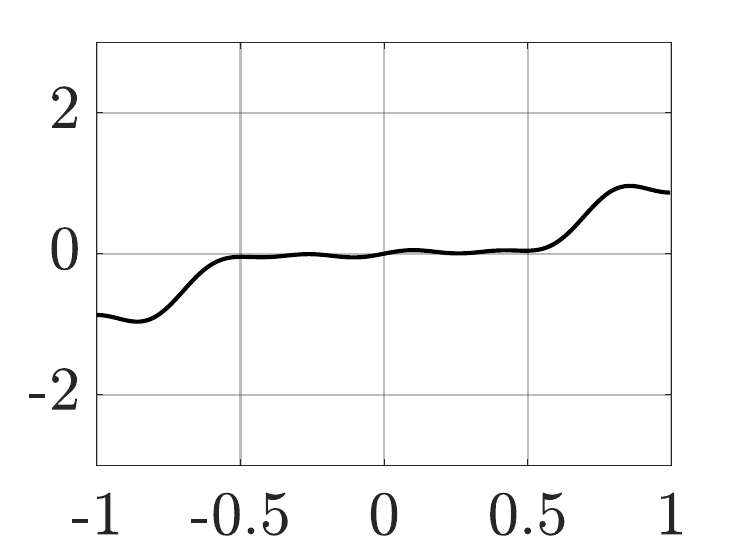}
         \caption[]{Hidden 5.} 
         \label{fig:act_lines_hidden5}
     \end{subfigure}
     \hspace{0pt}
     \begin{subfigure}[b]{0.32\columnwidth}
         \centering
         \includegraphics[width=\columnwidth]{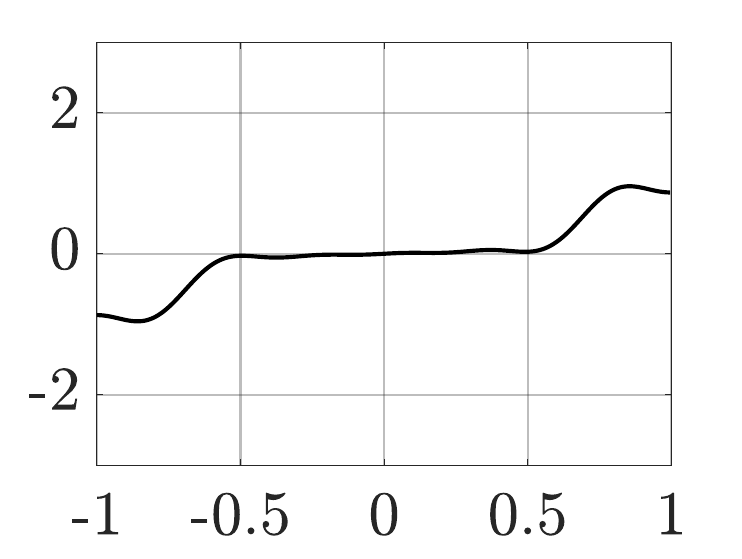}
         \caption[]{Hidden 6.} 
         \label{fig:act_lines_hidden6}
     \end{subfigure}
        \caption{\rev{AAF in the hidden layer for (P7).}}
        \label{fig:lines_aaf}
\end{figure}

\begin{figure}[t]
\centering
     \begin{subfigure}[b]{0.3\columnwidth}
         \includegraphics[width=\columnwidth]{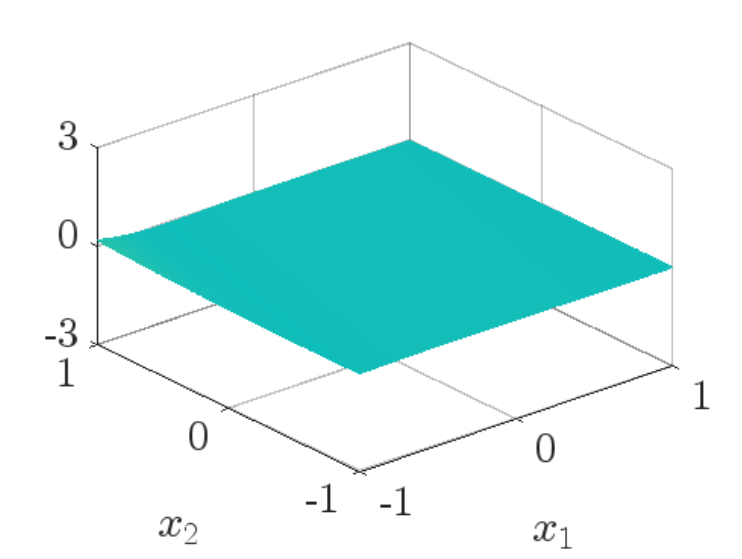}
         \caption[]{Hidden 1.} 
         \label{fig:bump_lines_hidden1}
     \end{subfigure}
     \hspace{0pt}
     \begin{subfigure}[b]{0.3\columnwidth}
         \includegraphics[width=\columnwidth]{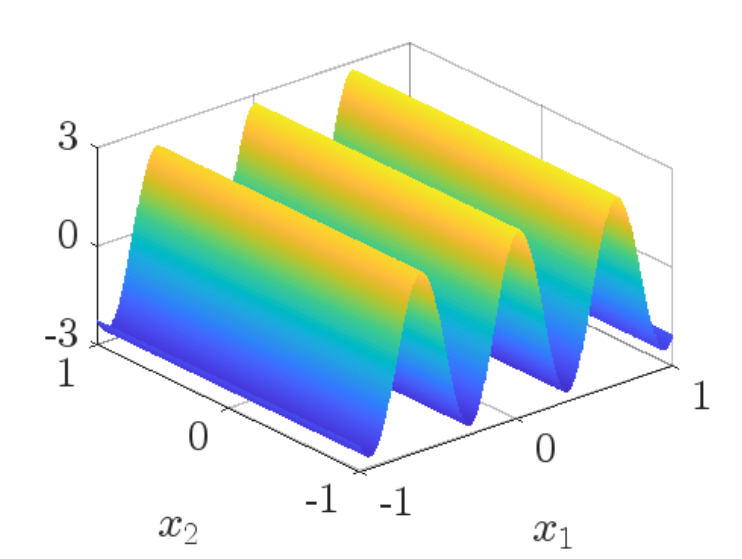}
         \caption[]{Hidden 2.} 
         \label{fig:bump_lines_hidden2}
     \end{subfigure}
     \hspace{0pt}
     \begin{subfigure}[b]{0.3\columnwidth}
         \includegraphics[width=\columnwidth]{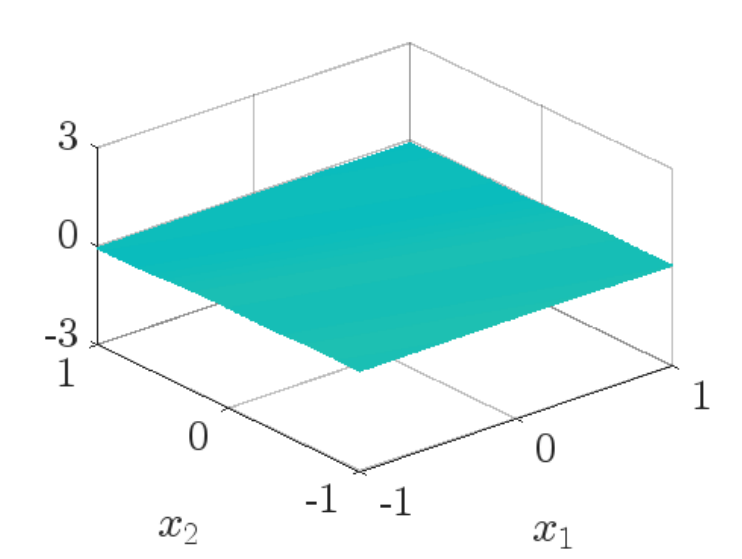}
         \caption[]{Hidden 3.} 
         \label{fig:bump_lines_hidden3}
     \end{subfigure}

     \vskip\baselineskip
     \begin{subfigure}[b]{0.3\columnwidth}
         \includegraphics[width=\textwidth]{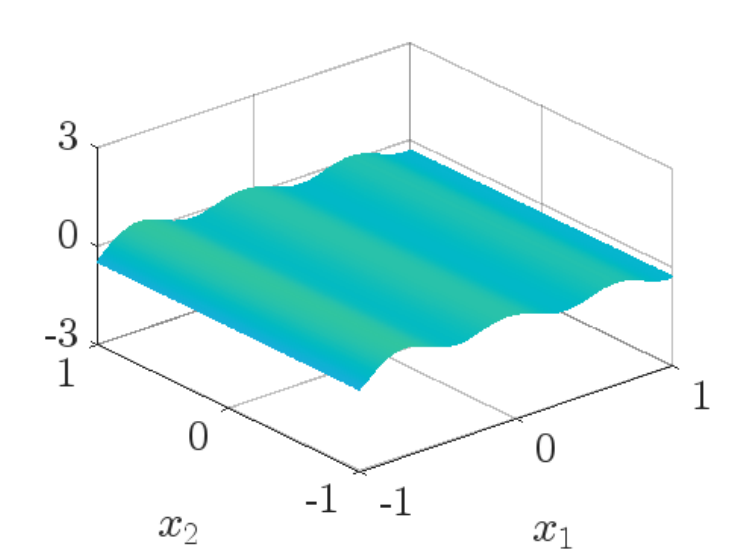}
         \caption[]{Hidden 4.} 
         \label{fig:bump_lines_hidden4}
     \end{subfigure}
     \hspace{0pt}
     \begin{subfigure}[b]{0.3\columnwidth}
         \centering
         \includegraphics[width=\columnwidth]{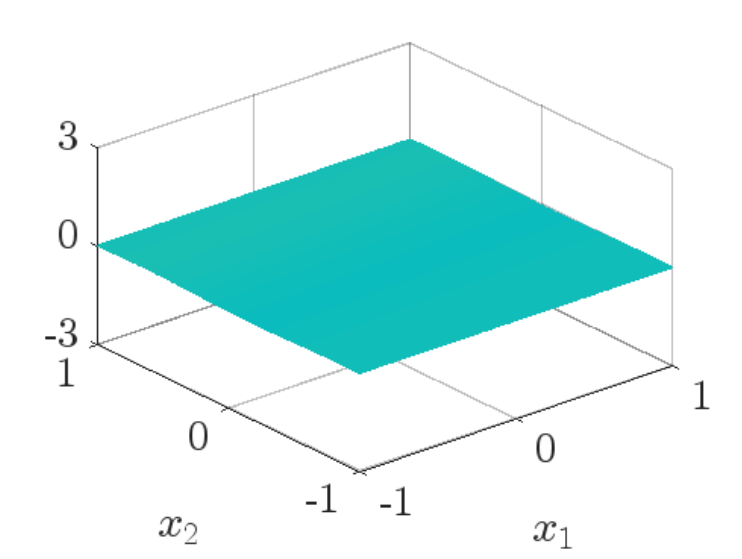}
         \caption[]{Hidden 5.} 
         \label{fig:bump_lines_hidden5}
     \end{subfigure}
     \hspace{0pt}
     \begin{subfigure}[b]{0.3\columnwidth}
         \centering
         \includegraphics[width=\columnwidth]{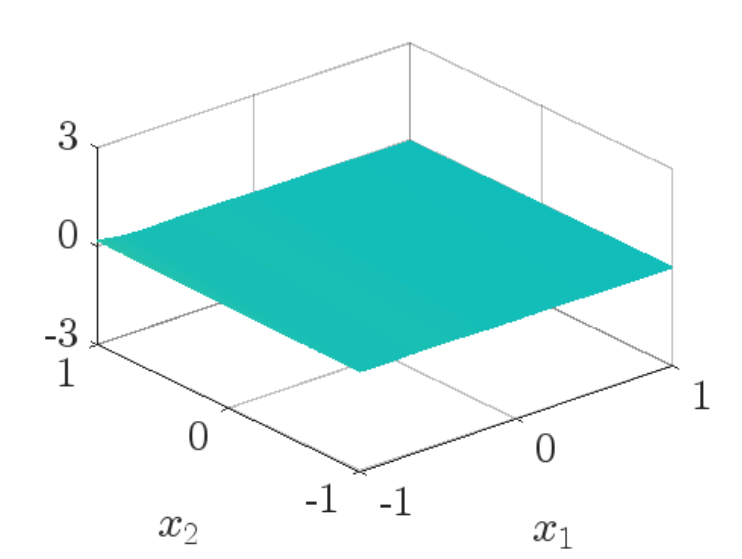}
         \caption[]{Hidden 6.} 
         \label{fig:bump_lines_hidden6}
     \end{subfigure}
        \caption{\rev{Bumps corresponding to the AAF in Fig. \ref{fig:lines_aaf}.}}
        \label{fig:lines_bump}
\end{figure}

\begin{figure}[h!]
\centering
     \begin{subfigure}[b]{0.49\columnwidth}
         \includegraphics[width=\columnwidth]{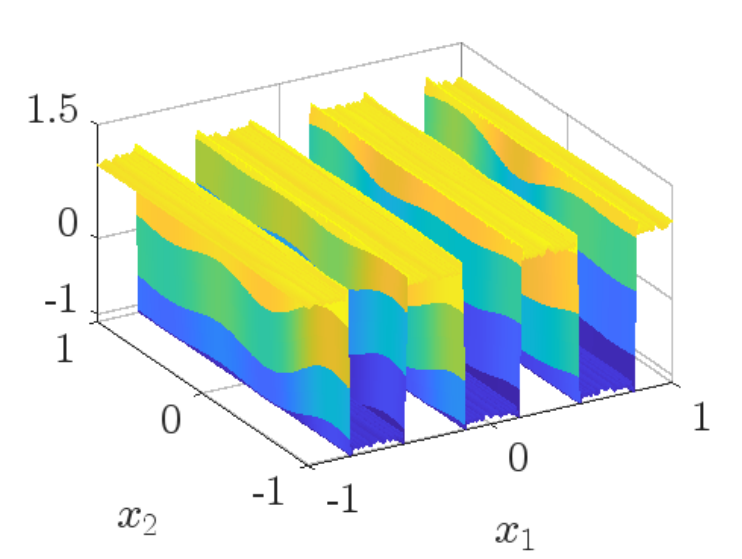}
         \caption[]{Global response for (P7).}
         \label{fig:response_lines}
     \end{subfigure}
     \hspace{0pt}
     \begin{subfigure}[b]{0.49\columnwidth}
         \includegraphics[width=\columnwidth]{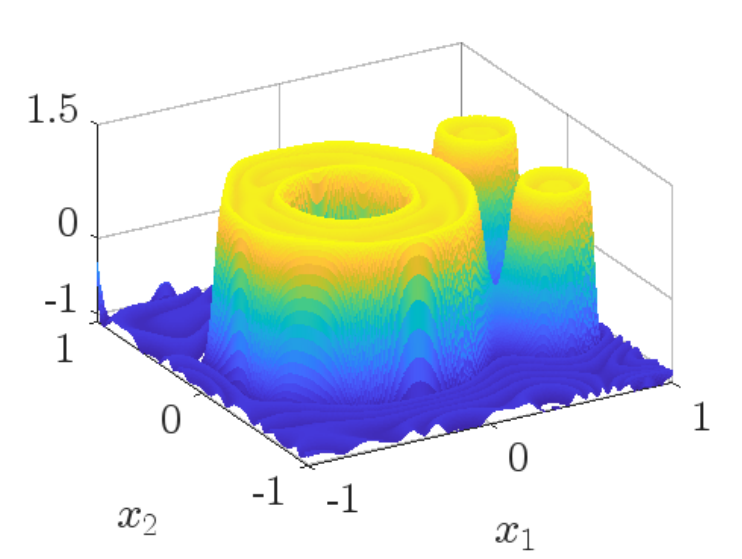}
         \caption[]{Global response for (P8).}
         \label{fig:response_face}
     \end{subfigure}
     \caption{\rev{Global response of the ENN for two classification problems.}}
     \label{fig:global_response}
\end{figure}

Fig. \ref{fig:lines_aaf} shows the AAF for (P7). \rev{Notice that only one of the activation functions has been adapted, namely Fig. \ref{fig:act_lines_hidden2}, which exploits the periodic nature of the DCT. The corresponding bumps are shown in Fig. \ref{fig:lines_bump}, where we clearly confirm that the other AAF have a dynamic range around $[-0.5,0.5]$, which results in zero bumps and, thus, in non-active neurons. The global response of the ENN is shown in Fig. \ref{fig:response_lines}. This is obtained by adding the bumps weighted by the corresponding linear weights of the output layer and, finally, applying the output AAF. As expected, the output AAF for classification problems is an approximation of the step function and it is shown in Fig. \ref{fig:act_out_classification}. Since there is only one neuron active, the global response corresponds to the step (or sign) function applied to the bump in Fig. \ref{fig:act_lines_hidden2}. Notice, however, that there is a flip between the bump and the global response, which happens because the corresponding linear weight of the output layer is negative.}
The fact of constraining the model to $Q=12$ coefficients makes it infeasible to generate an ideal step function in the output layer. Nevertheless, this solution is steeper than a standard sigmoid function, which reduces the error at data close to the boundary and increases the accuracy.
From this experiment we can also conclude that another advantage of ENN is that it allows to model the network width, this is, the required number of hidden neurons. In other experiments we found out that the ENN generates pairs of identical bumps but with opposite sign in the output linear weights. This means that the network neglects the information that comes from these two branches. See that this is impossible for the non-adaptive models, which have a very small accuracy even with 6 neurons.

\begin{figure}[t]
\centering
     \begin{subfigure}[b]{0.32\columnwidth}
         \includegraphics[width=\columnwidth]{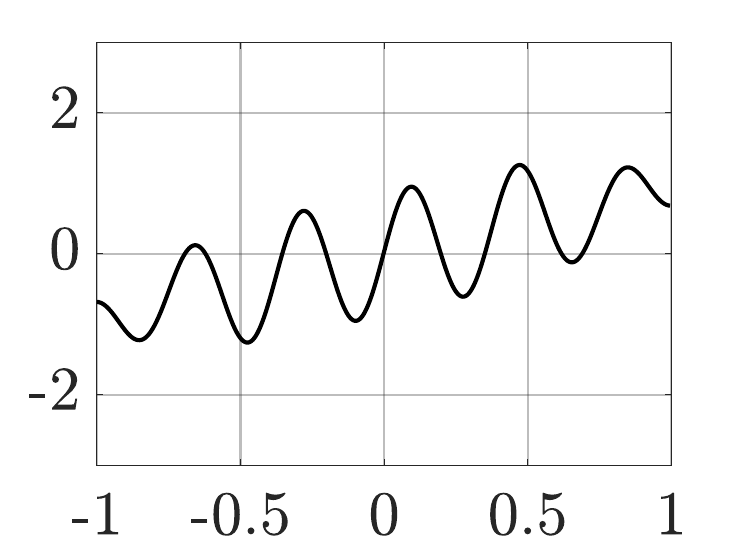}
         \caption[]{Hidden 1.} 
         \label{fig:act_face_hidden1}
     \end{subfigure}
     \hspace{0pt}
     \begin{subfigure}[b]{0.32\columnwidth}
         \includegraphics[width=\columnwidth]{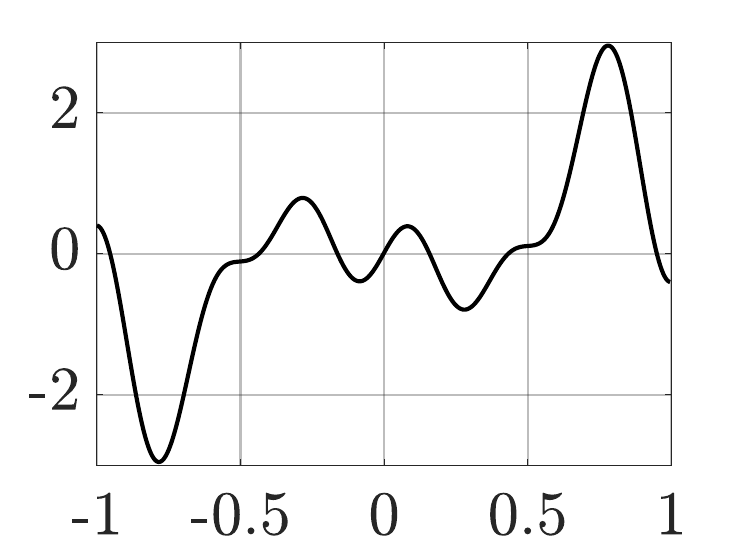}
         \caption[]{Hidden 2.} 
         \label{fig:act_face_hidden2}
     \end{subfigure}
     \hspace{0pt}
     \begin{subfigure}[b]{0.32\columnwidth}
         \includegraphics[width=\columnwidth]{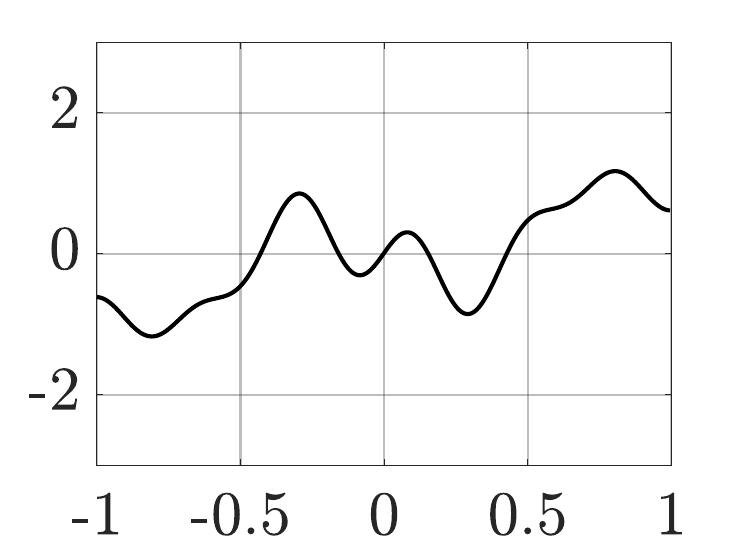}
         \caption[]{Hidden 3.} 
         \label{fig:act_face_hidden3}
     \end{subfigure}

     \vskip\baselineskip
     \begin{subfigure}[b]{0.32\columnwidth}
         \includegraphics[width=\textwidth]{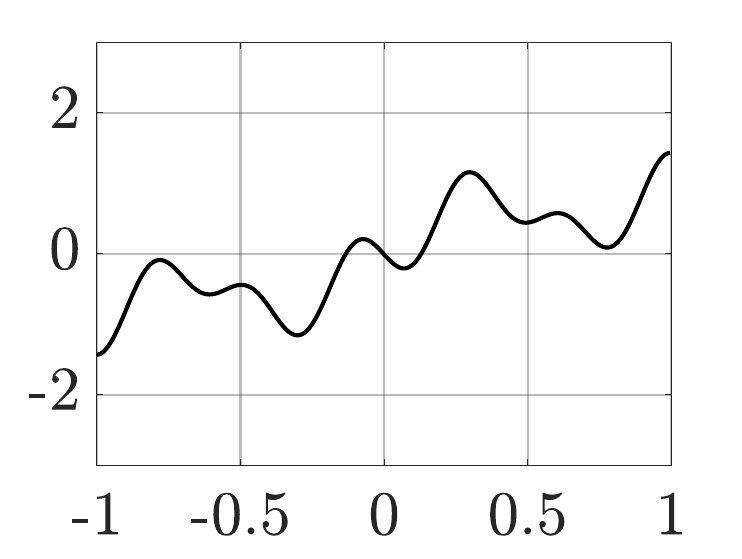}
         \caption[]{Hidden 4.} 
         \label{fig:act_face_hidden4}
     \end{subfigure}
     \hspace{0pt}
     \begin{subfigure}[b]{0.32\columnwidth}
         \centering
         \includegraphics[width=\columnwidth]{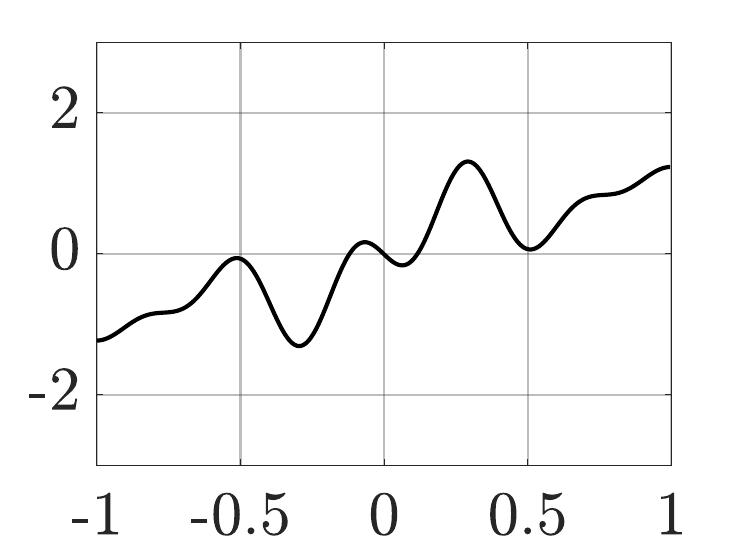}
         \caption[]{Hidden 5.} 
         \label{fig:act_face_hidden5}
     \end{subfigure}
     \hspace{0pt}
     \begin{subfigure}[b]{0.32\columnwidth}
         \centering
         \includegraphics[width=\columnwidth]{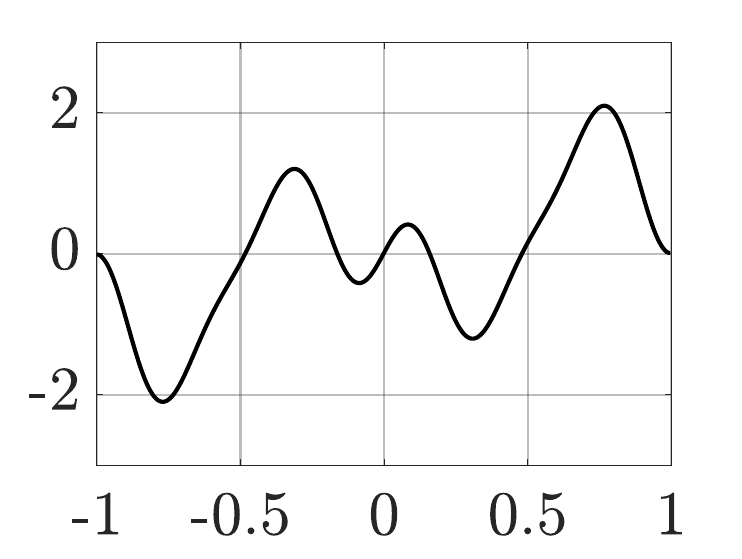}
         \caption[]{Hidden 6.} 
         \label{fig:act_face_hidden6}
     \end{subfigure}
        \caption{\rev{AAF in the hidden layer for (P8).}}
        \label{fig:face_aaf}
\end{figure}

\begin{figure}[t]
\centering
     \begin{subfigure}[b]{0.3\columnwidth}
         \includegraphics[width=\columnwidth]{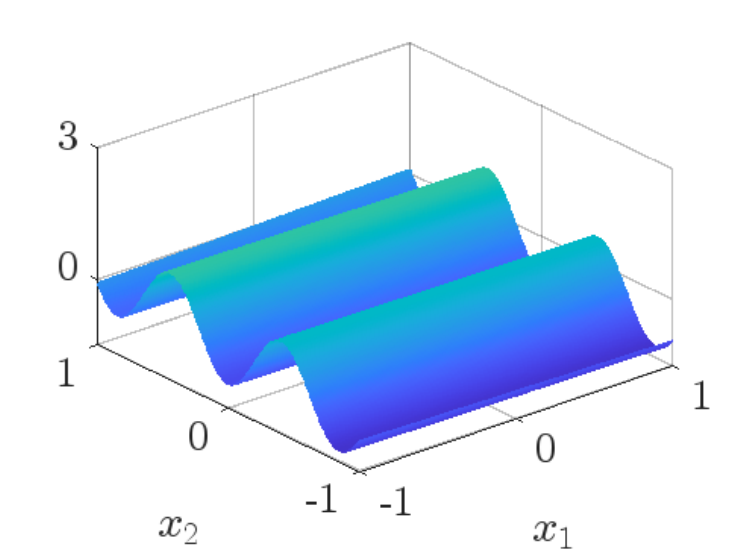}
         \caption[]{Hidden 1.} 
         \label{fig:bump_face_hidden1}
     \end{subfigure}
     \hspace{0pt}
     \begin{subfigure}[b]{0.3\columnwidth}
         \includegraphics[width=\columnwidth]{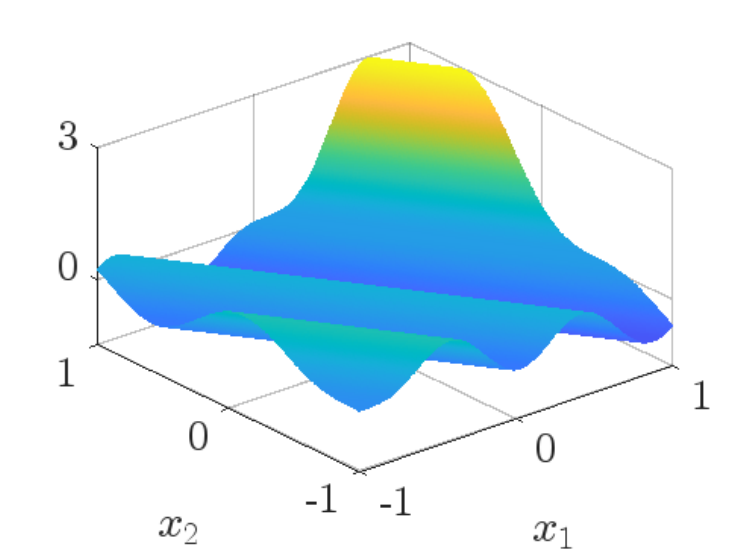}
         \caption[]{Hidden 2.} 
         \label{fig:bump_face_hidden2}
     \end{subfigure}
     \hspace{0pt}
     \begin{subfigure}[b]{0.3\columnwidth}
         \includegraphics[width=\columnwidth]{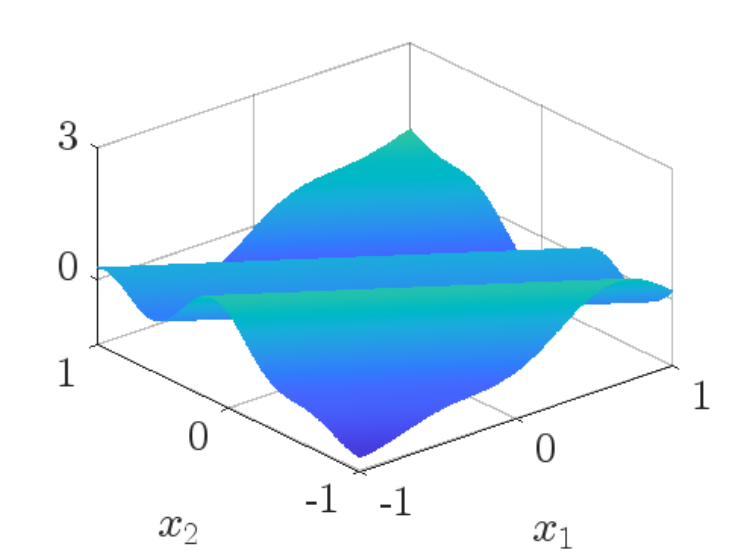}
         \caption[]{Hidden 3.} 
         \label{fig:bump_face_hidden3}
     \end{subfigure}

     \vskip\baselineskip
     \begin{subfigure}[b]{0.3\columnwidth}
         \includegraphics[width=\textwidth]{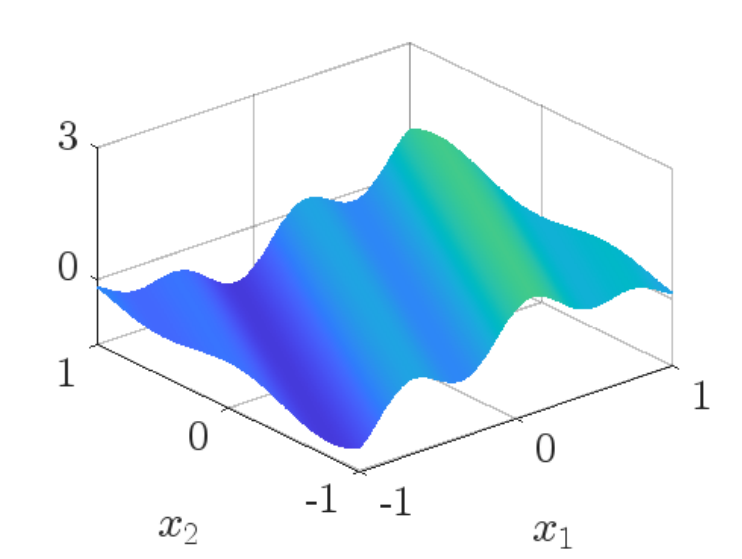}
         \caption[]{Hidden 4.} 
         \label{fig:bump_face_hidden4}
     \end{subfigure}
     \hspace{0pt}
     \begin{subfigure}[b]{0.3\columnwidth}
         \centering
         \includegraphics[width=\columnwidth]{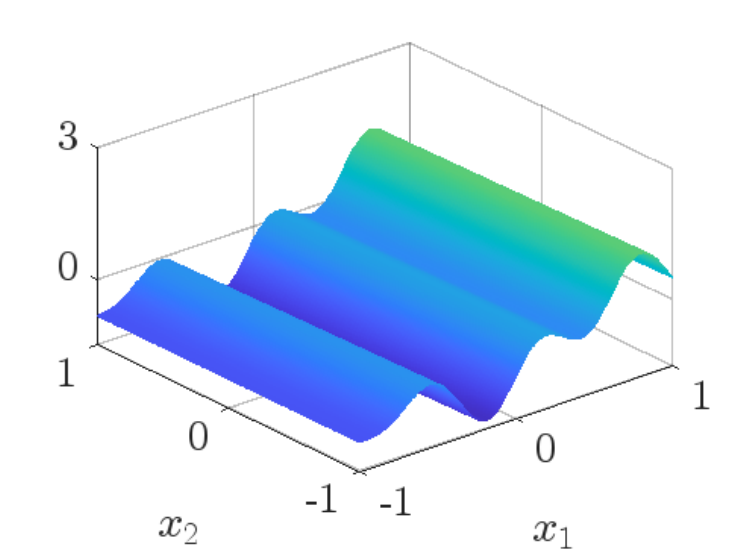}
         \caption[]{Hidden 5.} 
         \label{fig:bump_face_hidden5}
     \end{subfigure}
     \hspace{0pt}
     \begin{subfigure}[b]{0.3\columnwidth}
         \centering
         \includegraphics[width=\columnwidth]{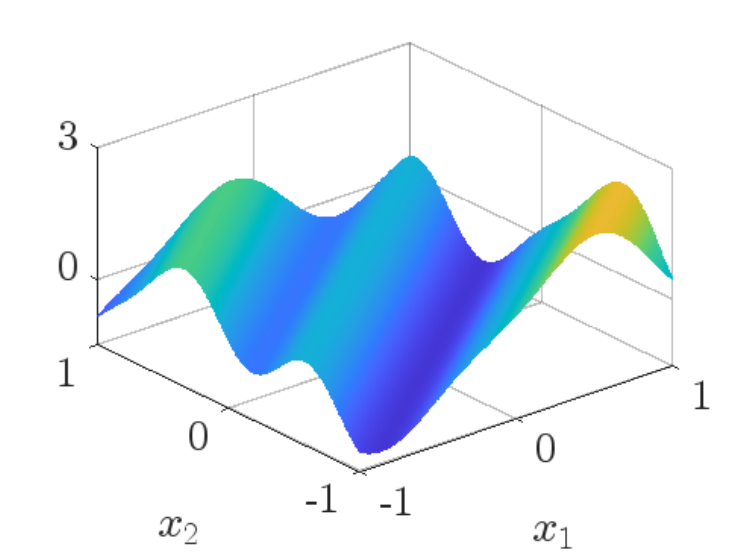}
         \caption[]{Hidden 6.} 
         \label{fig:bump_face_hidden6}
     \end{subfigure}
        \caption{\rev{Bumps corresponding to the AAF in Fig. \ref{fig:face_aaf}.}}
        \label{fig:face_bump}
\end{figure}

\begin{figure}[t]
\centering
     \begin{subfigure}[b]{0.23\columnwidth}
         \includegraphics[width=\columnwidth]{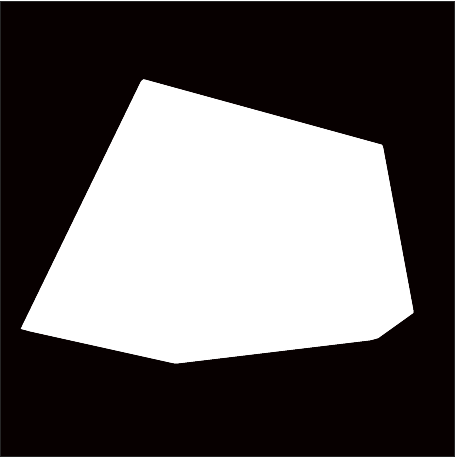}
         \caption[]{ReLU.} 
         \label{fig:map_face_RELU}
     \end{subfigure}
     \begin{subfigure}[b]{0.23\columnwidth}
         \includegraphics[width=\columnwidth]{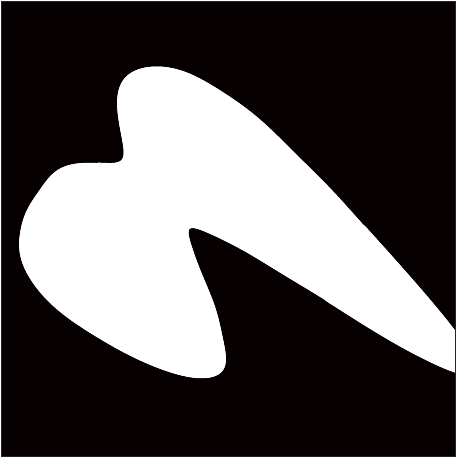}
         \caption[]{Sigm.} 
         \label{fig:map_face_Sigm}
     \end{subfigure}
     \hspace{0pt}
     \begin{subfigure}[b]{0.23\columnwidth}
         \includegraphics[width=\columnwidth]{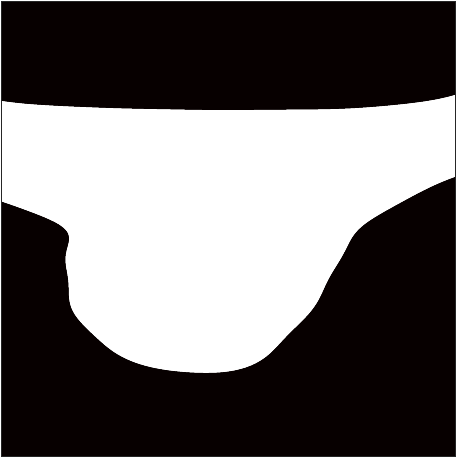}
         \caption[]{F-DCT.} 
         \label{fig:map_face_FDCT}
     \end{subfigure}
     \hspace{0pt}
     \begin{subfigure}[b]{0.23\columnwidth}
         \includegraphics[width=\columnwidth, angle=0]{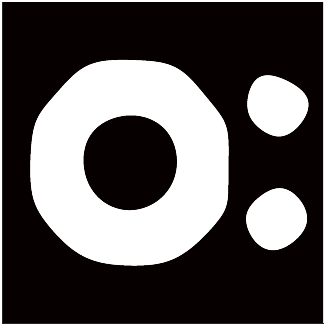}
         \caption[]{ENN.} 
         \label{fig:map_face_ADCT}
     \end{subfigure}

     \caption{\rev{Output decision map of (P8) for each of the models.}}
     \label{fig:face_maps}
\end{figure}

Fig. \ref{fig:face_aaf} shows the AAF for (P8), along with the corresponding bumps in Fig. \ref{fig:face_bump}. As expected, the ENN exploits the periodic nature of the DCT and different orientations to generate diversity.
Notice that there is almost no drop in accuracy between (P5) and (P8), although the latter is more complex. This is due to the expressiveness of the DCT, which allows to generate several peaks per bump. In other words, by allowing the activation to be non-monotone it reuses non-linearities to approximate the ring and both circles still with 6 neurons.
\rev{Fig. \ref{fig:response_face} shows the global response of the network} and Fig. \ref{fig:face_maps} shows the different decision boundaries for the different models. Despite the accuracy of the non-adaptive models in Table \ref{tab:classification_results} being around 80\%, the quality is very poor and only the ENN boundary recovers the ideal map.
The output activation function converges to the same configuration for all classification problems and it is shown in Fig. \ref{fig:act_out_classification}.

\subsection{Regression}
The network can also be trained for regression problems, this is, to approximate a function. Particularly, we compute bivariate scalar functions. Table \ref{tab:neuromorphic_MSE} shows the MSE achieved by the four different models for three different experiments.
\rev{While it may be impractical to implement a sum or product with a neural network as the model itself contains many sums and products, these experiments allow to test the learning capabilities of the ENN with respect to the benchmarks.}

\begin{figure}[t]
\centering
     \begin{subfigure}[b]{0.3\columnwidth}
         \includegraphics[width=\columnwidth]{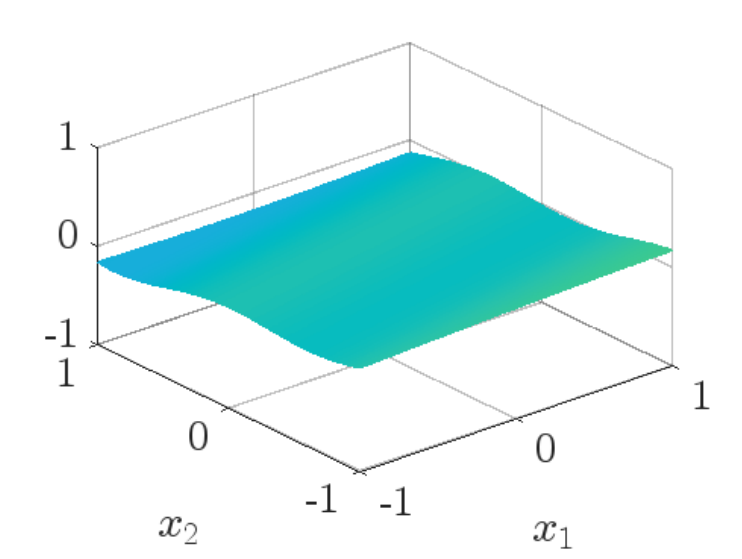}
         \caption[]{Hidden 1.} 
         \label{fig:bump_regression_hidden1}
     \end{subfigure}
     \hspace{0pt}
     \begin{subfigure}[b]{0.3\columnwidth}
         \includegraphics[width=\columnwidth]{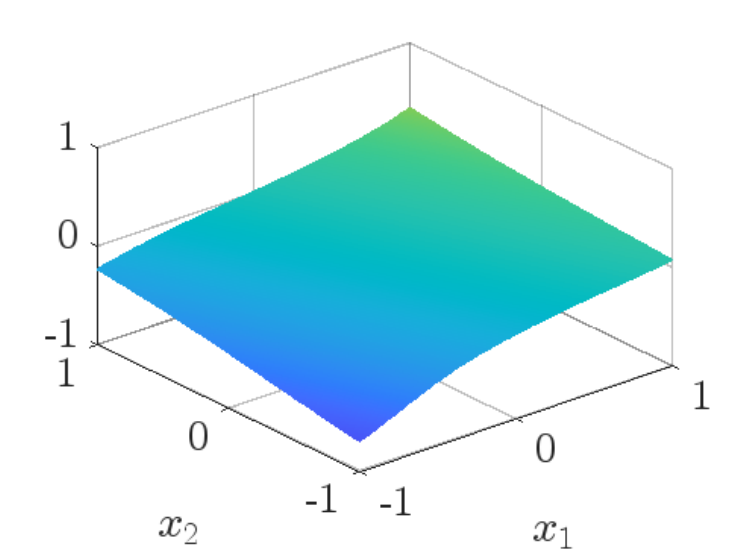}
         \caption[]{Hidden 2.} 
         \label{fig:bump_regression_hidden2}
     \end{subfigure}
     \hspace{0pt}
     \begin{subfigure}[b]{0.3\columnwidth}
         \includegraphics[width=\columnwidth]{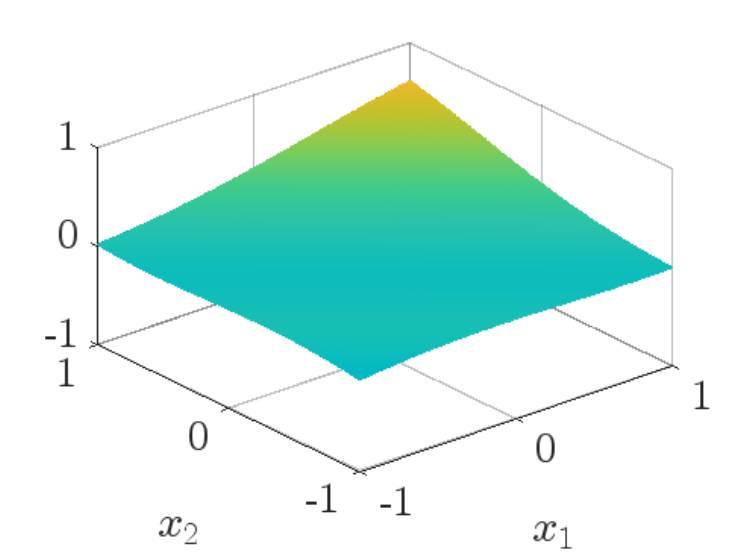}
         \caption[]{Hidden 3.} 
         \label{fig:bump_regression_hidden3}
     \end{subfigure}

     \vskip\baselineskip
     \begin{subfigure}[b]{0.3\columnwidth}
         \includegraphics[width=\textwidth]{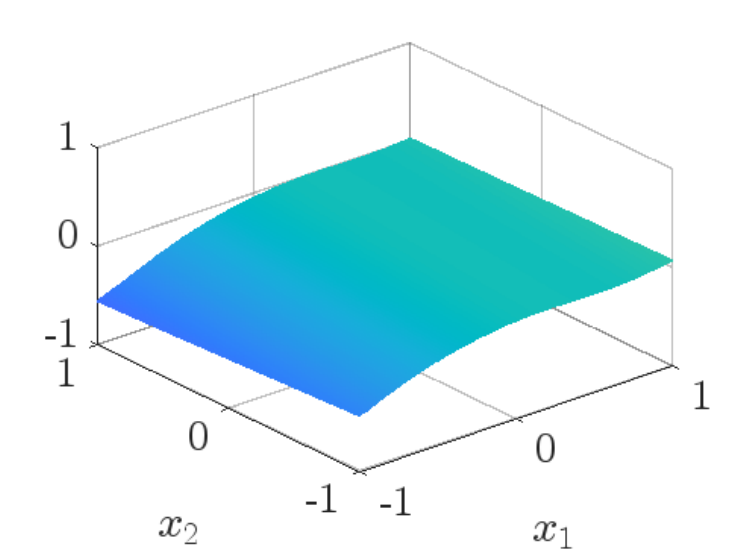}
         \caption[]{Hidden 4.} 
         \label{fig:bump_regression_hidden4}
     \end{subfigure}
     \hspace{0pt}
     \begin{subfigure}[b]{0.3\columnwidth}
         \centering
         \includegraphics[width=\columnwidth]{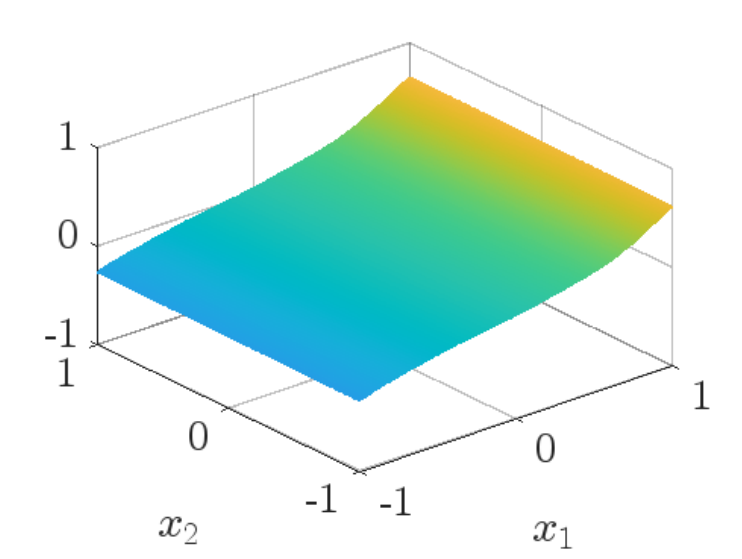}
         \caption[]{Hidden 5.} 
         \label{fig:bump_regression_hidden5}
     \end{subfigure}
     \hspace{0pt}
     \begin{subfigure}[b]{0.3\columnwidth}
         \centering
         \includegraphics[width=\columnwidth]{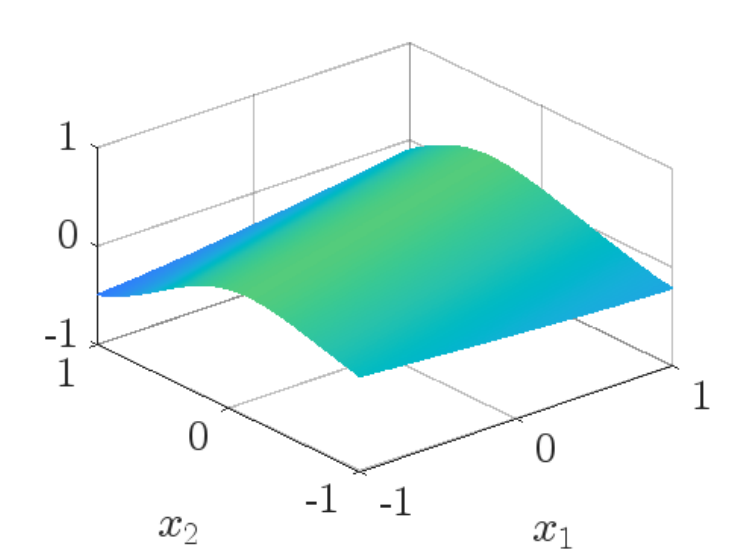}
         \caption[]{Hidden 6.} 
         \label{fig:bump_regression_hidden6}
     \end{subfigure}
        \caption{\rev{Bumps for the regression problem $(x_1^2+x_2^2)/2$.}}
        \label{fig:regression_bump}
\end{figure}

\rev{As expected, the gap in performance with respect to the benchmarks is larger in the second and third problem, since these are non-linear problems.}
Fig. \ref{fig:regression_bump} shows the different bumps for the regression problem $(x_1^2+x_2^2)/2$.
\rev{Since the function is symmetric around the center, the ENN chooses to work with non-periodic ranges of the non-linearities. In fact, all bumps correspond to flat surfaces that curve toward the edges. Recall that the curvature of the bump is undetermined as it also depends on the linear weights of the output layer. In fact, the weights of neurons 2, 4 and 6 are negative, meaning that all bumps are convex functions, as they need to be to generate a convex function.}

\begin{figure}[t]
\centering
     \begin{subfigure}[b]{0.49\columnwidth}
         \includegraphics[width=\columnwidth]{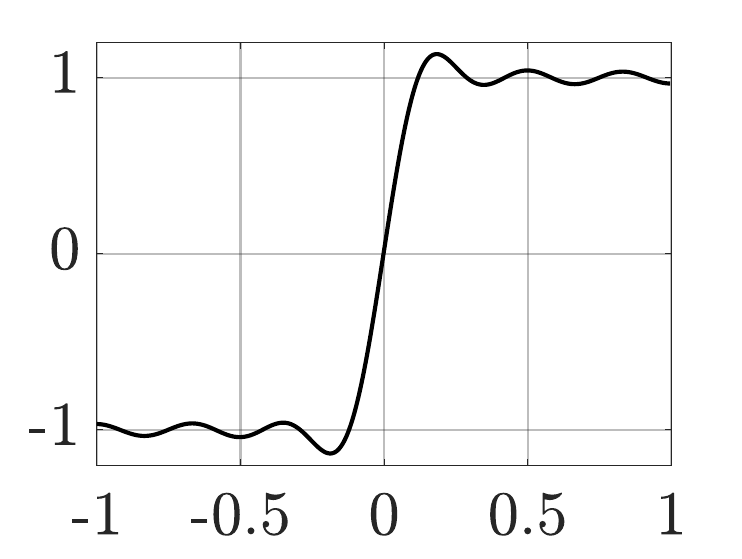}
         \caption[]{Classification.}
         \label{fig:act_out_classification}
     \end{subfigure}
     \hspace{0pt}
     \begin{subfigure}[b]{0.49\columnwidth}
         \includegraphics[width=\columnwidth]{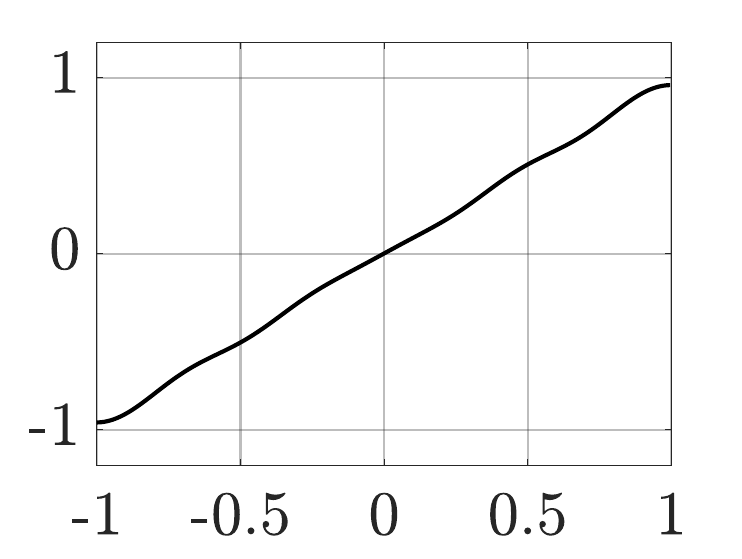}
         \caption[]{Regression.}
         \label{fig:act_out_regression}
     \end{subfigure}
     \caption{\rev{Output AAF for classification and regression.}}
     \label{fig:act_output}
\end{figure}

Regarding the output activation, it is adapted to the learning task and different from classification. As seen in Fig. \ref{fig:act_out_regression} the activation function approaches a linear function for all regression problems. The function is not perfectly linear because we constrain the system to $Q=12$ coefficients.
We underline the idea that in a non-adaptive setting one needs to choose the last non-linearity according to the task. This represents an improvement with respect to non-adaptive \rev{models} as it is general enough to encompass different tasks.

\begin{table}[t]
  \begin{center}
    \begin{tabular}{c|cccc}
      \toprule 
      Func. & ReLU & Sigm & F-DCT & ENN\\
      \midrule 
      $\frac{x_1+x_2}{2}$
      & ${2.9\cdot 10^{-3}}$ &
      ${2.1\cdot 10^{-2}}$ &
      ${3.5\cdot 10^{-3}}$ & 
      $\mathbf{2.3\cdot 10^{-6}}$\\
      
      $\frac{x_1^2+x_2^2}{4}$ &
      ${2.2\cdot 10^{-1}}$ &
      ${2.2\cdot 10^{-1}}$ & 
      ${2.2\cdot 10^{-1}}$ & 
      $\mathbf{9.5\cdot 10^{-6}}$\\
      
      $x_1x_2$ & 
      ${5.1\cdot 10^{-2}}$& 
      ${3.6\cdot 10^{-2}}$ & 
      ${1.6\cdot 10^{-1}}$ & 
      $\mathbf{7.6\cdot 10^{-5}}$\\
      \bottomrule
    \end{tabular}
    \caption{MSE in regression tasks.}
    \label{tab:neuromorphic_MSE}
  \end{center}
\end{table}


\subsection{Effect of the number of coefficients}
\begin{figure}[!ht]
    \centering
    \includegraphics[width=0.7\columnwidth]{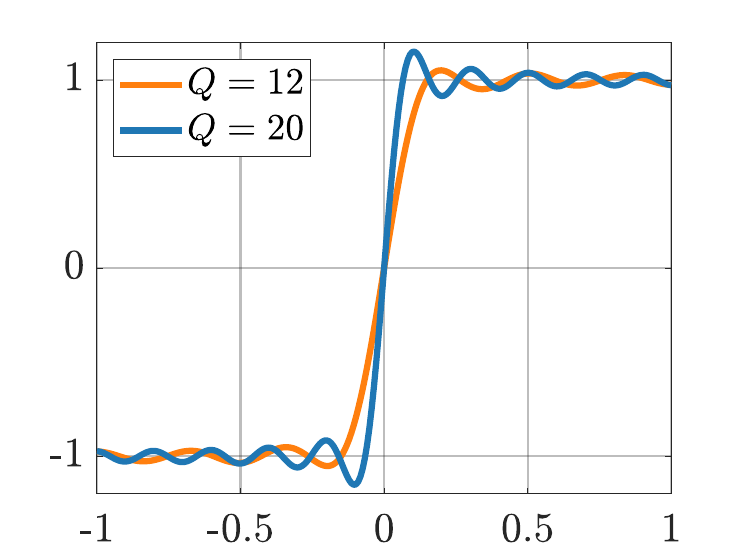}
  \caption{Output AAF in classification for different $Q$ values.}
  \label{fig:act_output_classification_different_Q}
\end{figure}

Increasing the number of coefficients $Q$ in the DCT representation theoretically provides more expressiveness to the network. However, in practice, increasing the number of trainable parameters hinders the convergence of the algorithm because there are more local minima. 
Fig. \ref{fig:act_output_classification_different_Q} shows the output AAF for $Q=\{12,20\}$. Increasing the number of coefficients allows to create a steeper transition, reducing the number of errors around the boundary. Doing so increases the Gibbs effect, although it produces no harm to the learning process because those samples are far from the boundary and always correctly classified as either 1 or -1.

\section{Conclusions}
In this paper we have presented ENN, a novel neural network \rev{model} with adaptive activation functions. Under a signal processing motivation we use the DCT to design the non-linear functions, whose coefficients can be trained using backpropagation. Specifically, the ENN is able to adapt the activation functions for classification and regression problems.
We provide insights in the interpretability of the network by recovering the concept of bump, this is, the response of each neuron in the output space. Through extensive experiments we determine that the expressiveness of a neural network highly depends on the activation function. Particularly, the key strength of ENN is the periodic nature of the DCT, providing high accuracy for a wide range of non-linear classification problems. In some cases outperforming state-of-the-art non-adaptive activation functions up to 40\% in accuracy.

\bibliographystyle{IEEEtran}
\bibliography{refs}

\end{document}